\RequirePackage{rotating} 
\documentclass[fleqn,usenatbib]{mnras}

\usepackage{caption}
\usepackage{subcaption}
\usepackage{graphicx}
\usepackage{array,multirow,graphicx}
 \usepackage{float}
 \usepackage{supertabular}
 \usepackage{amssymb}

\usepackage[T1]{fontenc}

\DeclareRobustCommand{\VAN}[3]{#2}
\let\VANthebibliography\thebibliography
\def\thebibliography{\DeclareRobustCommand{\VAN}[3]{##3}\VANthebibliography}

\usepackage{longtable}
\usepackage{graphicx}	
\def\code#1{\texttt{#1}}
\usepackage{newtxtext,newtxmath}

\title[Chemical Homogeneity and Ages of Stellar Strings]{A GALAH View of the Chemical Homogeneity and Ages of Stellar Strings Identified in Gaia}

\author[C. Manea et al.]{
Catherine Manea,$^{1}$\thanks{E-mail: cmanea@utexas.edu},
Keith Hawkins$^{1}$, and Zachary G. Maas$^{1,2}$
\\
$^{1}$Department of Astronomy, The University of Texas at Austin, 2515 Speedway Boulevard, Austin, TX 78712, USA \\
$^{2}$McDonald Observatory, The University of Texas at Austin, 2515 Speedway Boulevard, Austin, TX 78712, USA
}

\date{Accepted XXX. Received YYY; in original form ZZZ}

\pubyear{2021}

\begin{document}
 \label{firstpage}
\pagerange{\pageref{firstpage}--\pageref{lastpage}}
\maketitle

\begin{abstract} 
The advent of Gaia has led to the discovery of nearly 300 elongated stellar associations (called `strings') spanning hundreds of parsecs in length and mere tens of parsecs in width.  These newfound populations present an excellent laboratory for studying the assembly process of the Milky Way thin disk.  In this work, we use data from GALAH DR3 to investigate the chemical distributions and ages of 18 newfound stellar populations, 10 of which are strings and 8 of which are compact in morphology. We estimate the intrinsic abundance dispersions in [X/H] of each population and compare them with those of both their local fields and the open cluster M 67. We find that all but one of these groups are more chemically homogeneous than their local fields. Furthermore, half of the strings, namely Theias 139, 169, 216, 303, and 309, have intrinsic [X/H] dispersions that range between 0.01 and 0.07 dex in most elements, equivalent to those of many open clusters.  These results provide important new observational constraints on star formation and the chemical homogeneity of the local interstellar medium (ISM). We investigate each population's Li and chemical clock abundances (e.g., [Sc/Ba], [Ca/Ba], [Ti/Ba], and [Mg/Y]) and find that the ages suggested by chemistry generally support the isochronal ages in all but six structures.  This work highlights the unique advantages that chemistry holds in the study of kinematically-related stellar groups.
\end{abstract}




\section{Introduction}\label{sec:intro}
The chemodynamical nature of our Milky Way (MW) is a major topic of interest in modern astronomy .  Understanding the chemical, spatial, and dynamical structure of our MW not only informs our own Galaxy's formation and evolution but also offers a window into the physical processes that may govern the formation and evolution of other spiral galaxies. Stars are a major building block of the MW and can be used as a probe of its evolutionary history.  Young stars and stellar groups in particular are excellent tools for studying the structure and evolution of the MW's thin disk and the star formation processes that populate it.  We hereafter define a stellar `group' as a group of stars, either bound or unbound, that were born together of the same molecular cloud.  This definition includes open clusters, stellar associations, and other such populations of stars born together \citep[e.g.,][]{Trumpler1930, Lada2003, McKee2015, Krumholz2019}.

Various investigative tools can be applied to the study of young stellar groups to probe the formation and evolution of the thin disk.  The structure of the disk and spiral arms can be traced using the ages and kinematics of stellar groups.  Introducing chemistry into this process opens up an additional dimension of study, allowing one to probe the chemical structure, pollution history, and chemical homogeneity of the thin disk and its interstellar medium (ISM).  

In addition to probing the chemical and dynamical nature of the thin disk, young stellar groups are also effective tracers of the star formation process.  Most stars are born in groups within heirarchically-collapsing molecular clouds \citep[e.g.,][]{Portegies2010, Gieles2011, Bastian2012, Kruijssen2012, Krumholz2014}.  Molecular clouds are highly turbulent environments that homogenize the ISM to some degree prior to the onset of star formation, and simulations have shown that the degree of homogeneity depends on the initial homogeneity level of the ISM, the spatial scale, and other molecular cloud characteristics that influence mixing efficiency \citep[e.g.,][]{Feng2014, Armillotta2018, Krumholz2018}.  Because stars retain the chemical information of their birth environment like a fingerprint \citep[e.g.,][]{Feng2014}, we can observationally test the mixing processes within molecular clouds by studying the chemical homogeneity of birth groups.  In addition to tracing the mixing processes within star forming clouds, the intrinsic chemical dispersion of birth groups also informs the validity of strong chemical tagging, the practice of using stellar chemistry to identify field stars born of the same birth cloud \citep[e.g.,][]{Freeman2002, Blanco-Cuaresma2015, hawkins15, Hogg2016, PJ2020}.

There are various young products of star formation that can be used to study both the large-scale chemodynamical structure of the MW and the detailed mixing processes that occur in star forming molecular clouds.  Open clusters (OCs) are one such example.  OCs are massive ($\rm M\sim10^4-10^5~M_\odot$), gravitationally-bound stellar groups that are typically dispersed within a few hundred Myr to 1 Gyr due to interactions with the MW potential and various disk components, such as giant molecular clouds \citep[e.g.,][]{Krumholz2019}.  OCs are ideal for studying the chemodynamical structure and evolution of the MW due to their large stellar population sizes, which allow for greater precision in the determination of age (with uncertainties as low as $5\%$ from isochronal fitting) and mean chemical composition \citep[e.g.,][]{Friel1995, Santrich2018, Knudstrup2020, Anthony-Twarog2021}.  The chemical distributions of OCs have been used to trace the abundance gradients and pollution history of the ISM, along with the mixing efficiency of molecular clouds prior to star formation \citep[e.g.,][]{Yong2005, Reddy2012, Spina2021, Casamiquela2020}. Many OCs are observed to be highly chemically homogeneous, with dispersions in [X/H] and [X/Fe] ranging from 0.01 to 0.07 dex in various elements, including the $\rm \alpha$-elements, light-elements, and iron-peak elements \citep[e.g.,][]{Bovy2016a, Lambert2016, Ness2018, Kovalev2019, Poovelil2020}.  Some OCs are observed to be less homogeneous, but these instances are due to factors unassociated with the intrinsic homogeneity level of the primordial cloud.  These factors include atomic diffusion, non-local thermodynamic equilibrium effects, planetary engulfment, pollution by asymptotic giant branch (AGB) stars, and/or stochastic chemical 
enrichment, each of which alter the observed surface abundances of stars \citep[e.g.,][]{Blanco-Cuaresma2015, Liu2019, Souto2019, Casamiquela2020, Spina2021b}.

Though OCs are effective probes of the chemical and dynamical history of the MW and its star formation processes, they do not represent the most common star formation scenario \citep[e.g.,][]{Krumholz2019}. OCs may represent the upper limit of chemical homogeneity: because they are understood to form from massive molecular clouds with highly-efficient star formation, their mixing mechanisms may also be more efficient \citep[e.g.,][]{Larson1981, McKee2007, Kruijssen2012}. Thus, it is not clear whether we can apply the homogeneity results of OCs to other star formation scenarios \citep{Armillotta2018}.  It is therefore critical to explore chemical homogeneity in other contexts, such as gravitationally unbound stellar groups, the most common product of star formation \citep[e.g.,][]{Lada2003, McKee2007, Krumholz2019}.

The chemical homogeneity of lower-mass ($\rm M\sim12-10^4~M_\odot$), unbound stellar groups (also called associations) is lesser-studied relative to OCs (e.g., \citealp{Lada2003, Krumholz2019}).  Stellar associations are nearly impossible to recover kinematically beyond $\sim$100 Myr of age: because they are unbound, they diffuse rapidly and become phase-mixed with the disk on short timescales (\citealt{Krumholz2019}).   For this reason, most kinematically-confirmed stellar associations are young.  However, young groups can be difficult to study chemically because they often contain a plethora of hot (O, B, and A type) stars and stars with rapid rotation that possess broadened stellar lines that make abundance determinations difficult \citep{vanSaders2013, Soderblom2014}.  Despite these difficulties, studies have found that stellar associations tend to be chemically homogeneous either within measurement uncertainties \citep{DeSilva2007, Barenfeld2013} or down to 0.02 dex in [$\alpha$/Fe] and global metallicity, [M/H] \citep{Kos2020, Andrews2021}.  Additionally, a number of works have reported very low abundance dispersion in [X/Fe] vs. [Fe/H] among field disk stars (\citealp[e.g.,][]{Reddy2003}; \citealp{Bensby14}).

The advent of Gaia has allowed for the discovery of thousands of new stellar groups that allow for a diverse laboratory to study the star formation process and its influence on MW disk structure and evolution \citep[e.g.,][]{Faherty2018, Lee2018, Damiani2019, Pang2020, Tian2020, Kerr2021}.  In this work, we use the newfound kinematically-related stellar groups of \citet{Kounkel2019} to probe the star formation processes that occur in the Solar neighborhood.  \citet{Kounkel2019} (hereafter KC19) discovered nearly 300 extended stellar associations that are hundreds of parsecs in length and mere tens of parsecs in width, appropriately named `stellar strings' by the authors.  The isochronal ages of these strings range from $10^7$ to $10^9$~Myr, and the authors found that the older the string, the more diffuse and less numerous its stellar members.  KC19 suggest that they may be primordial in shape, as they share similar dimensions with giant molecular filaments, which are elongated molecular clouds that have been observed to trace spiral arms in our own MW and other spiral galaxies \citep[e.g.,][]{Ragan2014, Zucker2018}.  However, a chemodynamical investigation by \citet{Andrews2021} of Theia 456, one such string, may suggest that it was born  with a more compact morphology.

The majority of these strings lie either within the Local Arm or between the Local and the Sagittarius-Carina arms, and a few lie in the Sagittarius-Carina arm.  Thus, these strings provide an opportunity to study the star formation history of the Solar neighborhood and the star formation processes that build up local spiral arms.  In addition to these strings, KC19 discovered or recovered nearly 2,000 new or previously known associations and OCs.  We include several of these newfound, non-string-like, compact associations in our analysis, in addition to the strings.

In this work, we analyze the chemical abundances of ten newfound strings and eight newfound compact associations using data from GALAH DR3 \citep{GALAHDR3}.  We seek to address two major questions with our analysis:
\begin{enumerate}
    \item To what degree are these kinematically-related associations chemically related?
    \item What can the chemical distributions of these associations reveal about their formation mechanisms and birth environments?
\end{enumerate}
To facilitate the discussion of our results, we organize our paper in the following way. In Section \ref{Sec:Data}, we describe the discovery and selection of the structures explored in this work and the GALAH chemical data we use to study them. Sections \ref{Sec:Methods} and \ref{Sec:Results} present the methods used in and the results retrieved from our chemical analysis, which includes probing the intrinsic chemical homogeneity of each structure through Maximum Likelihood Estimation.  In addition, we report the absolute Li abundances and abundance ratios of four chemical clocks ([Sc/Ba], [Ca/Ba], [Ti/Ba], and [Mg/Y]) for each string, which enable us to chemically probe the youth and age of each structure.  In Section \ref{Sec:Discussion}, we place our results in the context of previous chemical investigations of stellar groups, and we discuss the possible formation scenarios for these strings given our homogeneity and age results.  In this section, we also propose a few avenues of follow up that will aid in further interpretation of the results.  We conclude with a summary in Section \ref{sec:conclusion}.

\section{Data}\label{Sec:Data}
\subsection{A catalog of Stellar Strings, Associations, and Open Clusters}\label{subsec:kc19}
In this study, we examine the chemical compositions of several newfound stellar groups discovered by KC19.  KC19 mined Gaia DR2 for stellar groups using Hierarchical Density-Based Spatial Clustering of Applications with Noise \citep[HDBSCAN,][]{HDBSCAN}, a clustering algorithm that groups high dimensional data by similarity.  HDBSCAN was ideal for their study because it does not require the number of output clusters to be specified; instead, it only requires the specification of a minimum number of clusters and a minimum size for each cluster.  KC19 focused their investigation on the Galactic midplane.  To prepare their HDBSCAN input data sample, they only considered Gaia sources within 30 degrees of the Galactic plane and restricted their parallaxes to greater than 1".  They made additional cuts on the data to remove stars with large astrometric uncertainties, which can cause poor performance in HDBSCAN.  After these cuts, 19.55 million stars within 1 kpc of the Sun remained and were input into HDBSCAN for clustering.  KC19 designated Galactic l and b positions, parallaxes, and proper motions as the input parameters on which HDBSCAN performed clustering.  KC19 required that each run of HDBSCAN output a minimum of 25 clusters with at least 40 stars each.

After performing HDBSCAN on the quality-cut Gaia DR2 data, KC19 found 1,901 kinematically-related stellar groups.  Of these stellar groups, 1,384 are newfound, and 284 of them possess an elongated morphology.  KC19 classify these elongated groups as `strings.'  Strings were self-consistently and visually classified, and KC19 note that visual classification was complicated by the fact that most structures are artificially elongated in \textit{l}.  KC19 also successfully recovered 198 pre-known clusters in their HDBSCAN run, validating their method.  KC19 emphasize that the structures they present are related solely through kinematics.  As such, they caution that several groups may be merely comoving and not true birth groups.  They estimate that up to 10\% of stars in these groups may be field contaminants that merely share the kinematics of the cluster with which they were associated.  

KC19 also estimate the ages of their retrieved groups using two methods: machine learning and isochrone fitting.  Their machine-learning approach involves creating a convolutional neural network (CNN), training it on OCs with well-determined ages, and applying it to their retrieved clusters to predict their ages.  KC19 report that this approach produced reasonable age estimates 44\% of the time, something that KC19 confirmed visually by comparing the isochrones constructed using the CNN-predicted parameters with the color-magnitude diagram of each group.  Their isochronal fitting approach produced successful fits to the color-magnitude diagrams of the groups 77\% of the time, something KC19 again confirmed visually.  KC19 estimate their reported ages to have uncertainties of $\sim$0.15 dex.  In this study, we are in a position to test the validity of these age estimates using a new axis, chemistry.

\subsection{GALAH DR3}\label{subsec:GALAHDR3}
In this work, we use data from the GALAH survey to examine the chemical compositions of the newfound stellar groups discovered by KC19.  GALAH systematically surveys the MW at low Galactic latitudes ($\rm |b| < 10^{\circ}$) selecting stars with $\rm 12 < V_{mag} < 14$ \citep{DeSilva2015}.  GALAH Data Release 3 (DR3) contains high-resolution (R $\sim$ 28, 000) spectra of 588,571 stars and reports stellar parameters such as effective temperature ($\rm T_{eff}$),  surface gravity (logg), and iron abundance ([Fe/H]) derived using Spectroscopy Made Easy (SME, \citealp{Valenti1996, Piskunov2017}) and 1D MARCS model atmospheres \citep{Gustafsson1975, Bell1976, Gustafsson2008}.  For each star, GALAH DR3 also reports its surface abundances of up to 30 elements: Li, C, O, Na, Mg, Al, Si, K, Ca, Sc, Ti, V, Cr, Mn, Co, Ni, Cu, Zn, Rb, Sr, Y, Zr, Mo, Ru, Ba, La, Ce, Nd, Sm, and Eu. The chemical space of GALAH is excellent for exploring the chemical natures of these stellar groups because it probes several different nucleosynthetic pathways and processes \citep[e.g.,][]{Burbidge1957, Wallerstein1997}, allowing access to a broad chemical profile for each star.

Throughout this study, we examine the elemental abundances of stars within each group and within the local field of each group.  To maximize the quality of the chemical data we consider, we employ the use of certain data cuts and flags.  We only consider GALAH data that satisfies the following requirements:

\begin{enumerate}
    \item \code{flag\_fe\_h} = 0
    \item 4000 K < \code{teff} < 6500 K
    and when considering individual elemental abundances:
    \item \code{flag\_X\_fe} = 0
\end{enumerate}

The first restriction ensures that the reported [Fe/H] abundance of each star, a crucial reference quantity for our homogeneity studies, is high-quality.  The second restriction eliminates very cool and very hot stars, where SME tends to fail.  The final restriction is only employed when considering the abundances of specific elements to ensure the reported abundances are high quality.  We also consider, and ultimately choose to not apply, the \code{flag\_sp} flag.  We conduct this study with and without restricting \code{flag\_sp} to 0 to determine the effect of this flag.  This flag is a general quality flag, and the various reasons this flag becomes non-zero include unreliable spectral broadening, poor S/N, spectral emission features, and poor Gaia astrometry \citep[See Table 6 in][for the full list]{Buder2018}.  We find that our results do not change significantly when employing this flag, though our sample size reduces slightly along with abundance dispersions.  The typical reasons for the activation of \code{flag\_sp} in our sample are signatures of potential binarity in the spectrum and t-SNE projected reduction issues or flux spikes in the spectra.  In an effort to present a meaningful analysis of each cluster in our pilot study, we prioritize a larger sample size over employing this final quality flag.

\subsection{Gaia eDR3 and Bailer-Jones et al. 2021 Distances}\label{subsec:eDR3}
In addition to chemical data from GALAH, we are interested in performing basic analyses of the 3-dimensional spatial distributions of the structures in our sample to understand their positional contexts in the Solar neighborhood.  We update both the KC19 and GALAH DR3 catalogs with Gaia eDR3 \citep{GaiaEDR3} astrometric and photometric data by performing a sky-position cross-match with a 1" search radius.

Stellar distance is an important quantity to constrain when determining 3-dimensional spatial distributions for our stellar groups, and a straightforward way to do this is through parallax inversion.  However, parallax inversion, when performed on stars with parallax uncertainties that are large ($ > 10\%$), can lead to biased distance estimates \citep{Astraatmadja2016} and large distance uncertainties ($^{+55}_{-45}$ pc at a distance of 500 pc).  Though the updated Gaia eDR3 catalog reduced random and systematic parallax errors by 30\% from the previous release, we wish to maximize the quality, precision, and accuracy of our distances.  Thus, we choose to adopt distances from the \citet{Bailer-Jones2021} catalog, which derived distances to stars sampled by Gaia eDR3 using a probabilistic approach.  The distances derived from this catalog use a distance prior that is based on a 3-dimensional model of the MW, which takes into consideration interstellar extinction and thus includes photometric information in the determination of stellar distance, along with astrometric information.  To ensure that we only consider high-quality photogeometric distances in our analysis, we restrict our analysis henceforth to stars with a photogeometric flag value 10000 $\leq$ \code{Flag} $\leq$ 10033 \citep[see][Table 2 for more information on this flag]{Bailer-Jones2021}.

\subsection{Final Sample}\label{finalsample}
Our final sample is constructed from the positional cross-match between the updated KC19 and GALAH DR3 catalogs using a search radius of 1".  To obtain a meaningful measure of chemical dispersion across each structure, we only consider structures that have at least 6 stellar members with at least 17 unflagged elemental abundances reported by GALAH.  These elements include O, Na, Mg, Al, Si, K, Ca, Sc, Ti, Cr, Mn, Fe, Ni, Cu, Zn, Y, and Ba.  We consider a sample size of 6 to be sufficient for this pilot study, and we draw the reader to works such as those of \citet{Ness2018} and \citet{Poovelil2020}, which drew meaningful conclusions about the chemical properties of clusters using similar sample sizes by using the particular statistical method of Maximum Likelihood Estimation (MLE), which we describe in detail in Section \ref{subsec:defininghomogeneity}.  We note that follow-up spectroscopic studies with greater spectral coverage of these groups would expand on the chemical homogeneity conclusions described in this work.

We find that 41 KC19 groups meet all of our GALAH sampling requirements, and 18 of these are newly discovered.  For the purposes of this study, we only consider newfound KC19 groups and leave the investigation of previously identified groups that are well-sampled by GALAH, such as Alpha Persei and Upper Sco, to future investigations (e.g., \citealt{Kos2020}).  Of the 18 newfound, well-sampled groups, 10 are strings and 8 are non-strings that have a compact, spherical morphology.  Table \ref{tab:structures} presents the average properties of these structures, and Table \ref{tab:targets} presents the high-quality (see Section \ref{subsec:GALAHDR3}), GALAH-sampled sources in each structure that we include in our analysis.  Across all stars in our final sample, the mean uncertainties in $\rm T_{eff}$, [Fe/H], and logg are 95 K, 0.08 dex, and 0.18 dex, respectively, and the mean signal to noise ratio per pixel is 35.

\begin{figure*}
	\includegraphics[width=2\columnwidth]{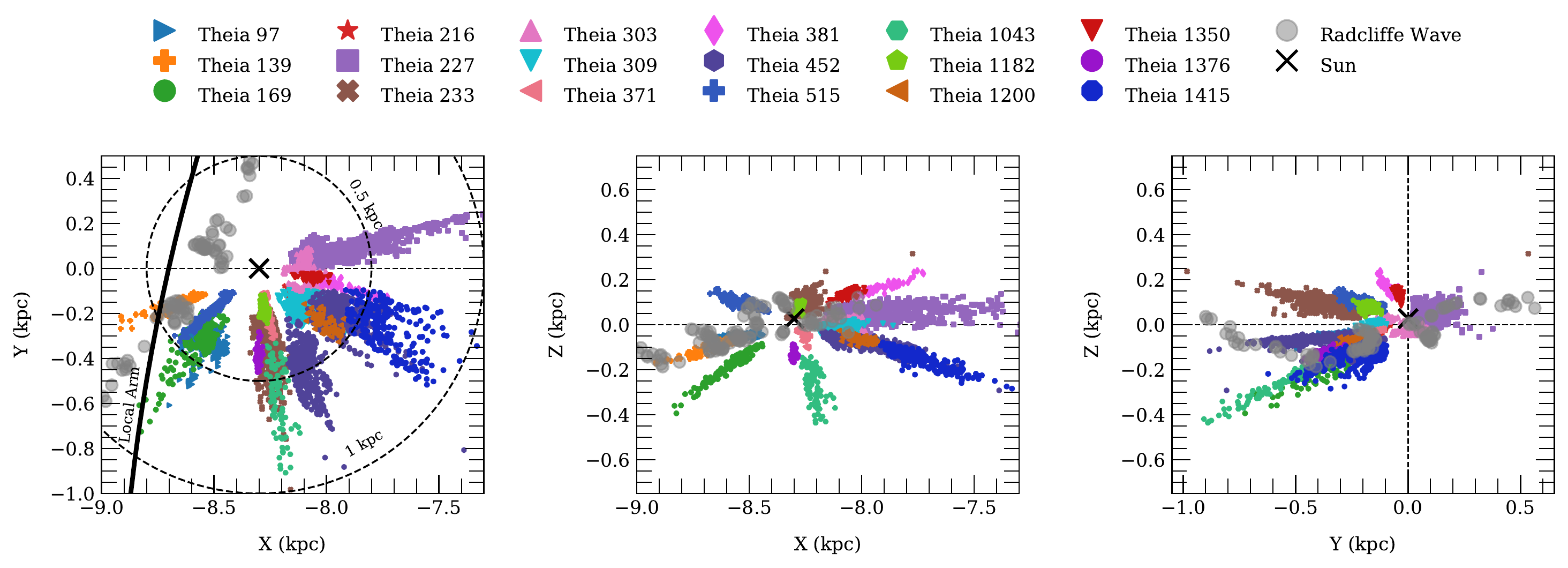}
    \caption{The 18 newfound structures investigated in this work, plotted in X-Y (left), X-Z (middle), and Y-Z (right) Galactocentric Cartesian coordinates.  The MW spiral arms model of \citet{Reid2014} (solid black curve) is included for reference.  The Radcliffe Wave is also included for reference (gray, \citealp{RadcliffeWave}).  Theia 139 directly intersects the Radcliffe wave, and Theias 97, 169, and 515 appear to lie in the Local arm, while the remaining, older structures reside in the gap between the Local and Sagittarius-Carina arms.}
    \label{fig:skyplot}
\end{figure*}

In this work, we pair our examination of the chemical distributions of each of the 18 newfound, well-sampled KC19 groups with basic analyses of their 3-dimensional Galactocentric positions to understand where these structures reside in the Solar neighborhood.  To do this, we determine the 3-dimensional Galactocentric-Cartesian coordinates of the stellar members of each structure using their Gaia eDR3 right ascension and declination, distances adopted from \citet{Bailer-Jones2021}, and Astropy's \code{SkyCoord} class\footnote{We assume a Solar position of Galactocentric-Cartesian X, Y, Z = (8.300, 0.000, 0.027) kpc \citep[adopted from][]{Gillessen2009, Chen2001} when utilizing \code{SkyCoord}.} \citep{Astropy}.  Figure \ref{fig:skyplot} presents the spatial distributions of each structure in 3-dimensional Galactocentric Cartesian coordinates, with each panel displaying the view in the X-Y (left panel), X-Z (center panel), and Y-Z plane (right panel).  Four structures in our sample, Theias 97, 139, 169, and 515, lie near the Local Arm and the Radcliffe Wave, an undulating wave of gas composed of interconnected star forming clouds \citep{RadcliffeWave}.  The remaining structures reside in the inter-arm region between the Local and Sagittarius-Carina arms.  Figure \ref{fig:CMDs} presents the color-magnitude diagrams of the structures in our sample.  Absolute G magnitudes are derived using Gaia eDR3 \code{phot\_g\_mean\_mag}, distances adopted from \citet{Bailer-Jones2021}, and dust extinction values adopted from \citet{GaiaEDR3}.  For each structure's color-magnitude diagram, we also present PARSEC isochrones \citep{PARSEC} associated with the ages estimated by KC19, which range from 33 Myr (Theia 97) to 1.72 Gyr (Theia 1415).

\subsubsection{Local Field Cylinders for Each Structure}\label{subsec:localfield}
\begin{figure*}
	\includegraphics[width=2\columnwidth]{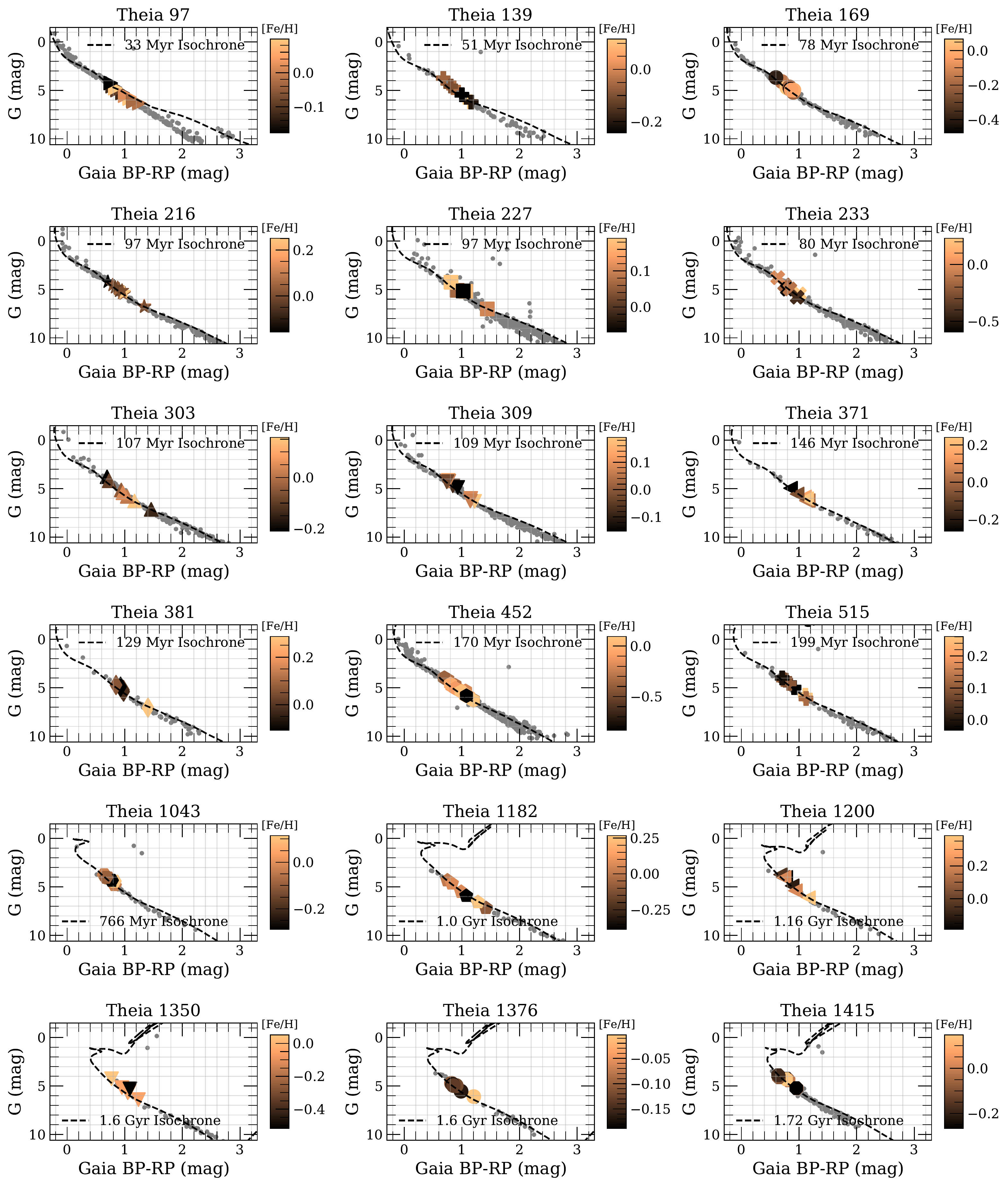}
    \caption{Color-magnitude diagrams of the structures investigated in this work, with G-band extinction values adopted from \citet{GaiaDR2}.  Grey points represent the members comprising structure.  Symbols represent high-quality GALAH-sampled members and are colored by [Fe/H].  The dashed line is the PARSEC isochrone of the best-fit age as determined by KC19.  All structures have tight CMDs, supporting potential coevality.}
    \label{fig:CMDs}
\end{figure*}

In this study, we compare the chemical distribution of each structure to nearby stars to determine how chemically distinguishable stellar groups are from their local field.  We choose to define the local field of each structure as all FGK dwarfs that lie within a bounding cylinder of length 500 pc and radius 70 pc that is coaxial with the length axis of the structure and centered on the structure's 3-dimensional centroid.  This geometry ensures that our local field is as small as possible while still encompassing each structure, and this cylindrical shape assists in ensuring that our measurements of abundance dispersion are not dominated by an excessively large local field population.  We restrict our analysis of each local field to FGK dwarfs because the GALAH-sampled stars in each group that we consider are exclusively FGK dwarfs due to 1) the temperature cuts we employ, and 2) the young natures of each structure, as suggested by their color-magnitude diagrams (Figure \ref{fig:CMDs}).  To define each structure's local field population, we begin by determining the 3-dimensional centroid of each structure by calculating the mean Galactocentric X-, Y-, and Z-coordinates of the members of each structure.  Next, we determine the length axis of each structure by applying the principal component analysis (PCA) function in Scikit-Learn's \code{decomposition} package \citep{scikit} to the Galactocentric X-, Y-, and Z-coordinates of each structure, where we take the first principal component to be the length axis.  We then determine the Galactocentric-Cartesian coordinates of the full high-quality GALAH catalog using the same method used to determine Galactocentric-Cartesian coordinates for each KC19 structure (see Section \ref{finalsample}).  We only consider high-quality (as defined in Section \ref{subsec:GALAHDR3}) GALAH sources that lie within 70 pc of the length axis of each structure and, when projected onto the length axis, lie within 250 pc of the structure's centroid.  Finally, to ensure we are only sampling FGK dwarfs, we restrict our local field cylinders to stars with GALAH \code{logg} greater than 3.5 and absolute G magnitude greater than 2.0 mag.

\subsubsection{Comparing to the Chemical Dispersion of Well-Studied Open Cluster M 67}\label{subsec:m67_method}
In addition to comparing the chemical distributions of each structure to those of their local fields, we also compare them to that of M 67.  M 67 is an OC whose chemical homogeneity has been extensively studied. \citet{Bovy2016a}, for example, found that M 67 is homogeneous in C, N, O, Na, Mg, Al, Si, S, K, Ca, Ti, V, Mn, Fe, and Ni at the 0.01 to 0.07 dex level, while \citet{Liu2016} and \citet{Liu2019} found that subgiants in M 67 are homogeneous at the 0.02 dex level in up to 22 elements.  \citet{Gao2018} studied M 67 using GALAH data and found that M 67 has dispersions in abundances between 0.03 and 0.07 dex in [Fe/H], [Na/Fe], [Mg/Fe], [Al/Fe], and [Si/Fe] for turn-off, subgiant, and giant stars in M 67.  Alongside reports of homogeneity among stars in M 67, works also note that there are abundance differences of up to 0.3 dex between stars in different evolutionary states (i.e. turn-off, subgiant, and giant stars) in M 67, possibly due to atomic diffusion effects \citep{Liu2016, Souto2018, Liu2019}.  Interestingly, \citet{Gao2018} note that abundance differences between stars of differing evolutionary state in M 67 are significantly lessened when non-LTE is assumed.

The purpose of this comparison is to examine how the chemical distributions of the KC19 structures compare to a well-studied OC that has been observed with the same setup and thus that suffers from similar systematics.  We thus choose to use the M 67 membership catalog of \citet{Gao2018} to select stars for this comparison and only consider turn-off stars.  We select only turn-off stars from the M 67 \citet{Gao2018} catalog because they are the closest in evolutionary state to the dwarfs sampled by GALAH in our structures.  The purpose of this work is not to re-examine the chemical homogeneity of M 67 but rather to use it as a comparison object with which we can contextualize the homogeneity of the KC19 structures.  To extract the GALAH DR3 chemical data for the turn-off stars in M 67, we cross match the M 67 membership catalog of \citet{Gao2018} with the quality-cut GALAH DR3 catalog (see Section \ref{subsec:GALAHDR3}) using turn-off member stars' \code{sobjectid}.  This results in 41 stars that have a mean $\rm T_{eff}$,  logg, [Fe/H], and signal to noise ratio per pixel of $\sim$5600 $\pm$ 500 K, 3.73 $\pm$ 0.6 dex, -0.05 $\pm$ 0.05 dex, and 71 $\pm$ 46, respectively.

\begin{table}
	\centering
	\caption{General properties of the 18 newfound stellar structures investigated in this work.  All columns are taken from \citet{Kounkel2019} with the exception of the aspect ratio (AR) column, which reports a metric that compares the length of each string to its width, as is customary when describing molecular filaments of similar shapes and sizes \citep[e.g.,][]{Ragan2014, Zucker2018}, and the N$_{\rm G}$ column, which presents the number of stars with high-quality GALAH-reported abundances, as defined in Section \ref{subsec:GALAHDR3}.  Note that the column titled `N$_{\rm *}$' reports the total number of stars comprising the structure, as determined by \citet{Kounkel2019}.} 
	\label{tab:structures}
	\begin{tabular}{p{.4 cm}p{.3cm}p{1.0 cm}p{.6 cm}p{.7 cm}p{.7 cm}p{.7 cm}p{.3 cm}p{.3 cm}} 
		\hline
		Theia & N$_{\rm *}$ & log$_{10}$(Age) & String? & Width & Height & Length & AR & N$_{\rm G}$\\
		  &   &   & (y/n) & (pc) & (pc) & (pc) & & \\
		\hline
		97 & 977 & 7.52 & y & 53 & 8 & 193 & 4:1 &  13\\
        139 & 129 & 7.71 & y & 14 & 10 & 196 & 15:1 &  10\\
        169 & 284 & 7.90 & y & 38 & 8 & 181 & 5:1 &  10\\
        216 & 441 & 7.99 & y & 14 & 7 & 88 & 6:1 &  14\\
        227 & 999 & 7.99 & n & 17 & 32 &  &  & 7\\
        233 & 618 & 7.90 & y & 63 & 18 & 251 & 4:1 & 7\\
        303 & 335 & 8.03 & y & 64 & 8 & 284 & 5:1 & 8\\
        309 & 582 & 8.04 & y & 38 & 11 & 193 & 5:1 & 13\\
        371 & 86 & 8.17 & n & 8 & 8 &  &  & 9\\
        381 & 81 & 8.11 & n & 14 & 9 &  &  & 8\\
        452 & 999 & 8.23 & y & 142 & 11 & 486 & 3:1 & 19\\
        515 & 178 & 8.30 & n & 10 & 12 &  &  & 7\\
        1043 & 93 & 8.88 & n & 27 & 10 &  &  & 9\\
        1182 & 92 & 9.00 & n & 10 & 15 &  &  & 7\\
        1200 & 69 & 9.06 & y & 23 & 7 & 142 & 6:1 & 8\\
        1350 & 58 & 9.20 & n & 10 & 10 &  &  & 10\\
        1376 & 53 & 9.20 & n & 8 & 11 &  &  & 8 \\
        1415 & 264 & 9.23 & y & 72 & 15 & 351 & 5:1 & 10\\
		\hline
	\end{tabular}
\end{table}

\begin{table*}
	\centering
	\caption{The final sample of stars included in our analysis.  For each structure, we only consider GALAH-sampled sources that fulfill to the GALAH flag and temperature requirements detailed in Section \ref{subsec:GALAHDR3}. Full dataset available as a supplementary table in machine-readable format.}
	\label{tab:targets}
	\begin{tabular}{ccccc} 
		\hline
		Theia & Gaia DR2 ID & GALAH \code{sobject\_id} & RAJ2000 & DEJ2000 \\
		 &  &  & (deg) & (deg) \\
		\hline
97 & 2944871852953405952 & 170106003601353 & 95.37812 & -16.407782 \\
97 & 5607168526072940288 & 150406001401095 & 105.37533 & -31.457504 \\
97 & 5602641317006928384 & 160130004101305 & 107.36293 & -34.604156 \\
97 & 2944699916821425152 & 170106003601292 & 94.70931 & -16.303555 \\
97 & 2944687547315653248 & 170106003601269 & 94.592865 & -16.515806 \\
. . . & . . . & . . . & . . . & . . . \\
1415 & 5920368291595355392 & 160518002901290 & 266.2051 & -55.5544 \\
1415 & 5913040291426212224 & 150830002301211 & 261.5889 & -58.820694 \\
1415 & 5918660170225603200 & 150703004101311 & 266.63803 & -57.375683 \\
1415 & 5909156919425906304 & 140809002101360 & 267.29892 & -64.55563 \\
1415 & 6703479544222461184 & 160815003101208 & 275.64688 & -49.325153 \\
		\hline
	\end{tabular}
\end{table*}

\section{Methods}\label{Sec:Methods}
\subsection{Defining and Estimating Intrinsic Homogeneity}\label{subsec:defininghomogeneity}
A primary goal of this investigation is to constrain the chemical natures of 18 newfound stellar groups, and one way we do this is by examining the chemical homogeneity of each structure.  We use the terms homogeneity and dispersion interchangeably; a structure that is highly homogeneous in [X/H] has a low dispersion in [X/H] across its constituent stars.  If we assume that the [X/H] abundance distribution of each structure can be modeled as a Gaussian, then the standard deviation of the distribution is a measure of the structure's abundance dispersion.  However, in practice, this standard deviation is a convolution of the \textit{intrinsic} abundance dispersion and the abundance uncertainties.  In this work, we are particularly interested in the \textit{intrinsic}, underlying abundance dispersions of the KC19 structures.  We can directly compare our measurements of intrinsic dispersion to those determined by other observational and theoretical works to contextualize our results in the landscape of the MW's star formation integrants (ISM, molecular clouds) and products (stellar birth groups).

Some studies of OCs have estimated intrinsic dispersions in chemical abundance by minimizing uncertainties in measured abundance by using high-resolution, high signal to noise ratio (S/N) spectral data (e.g. \citealp{Lambert2016}, who obtained a spectral resolution of $\rm R\sim55,000$ and S/N$\sim 100-190$, achieving abundance uncertainties as low as 0.01 in [X/Fe]).  Others have estimated upper limits on the intrinsic abundance dispersion of a cluster by measuring the root-mean square star-to-star scatter in abundance \citep[e.g.,][]{DeSilva2007}.  Many studies of OC and globular cluster homogeneity estimate intrinsic abundance dispersion through line-by-line differential abundance analysis on solar twins, stars with similar atmospheric properties to the Sun, in an effort to avoid uncertainties in abundances due to evolutionary stage, metallicity, and systematics \citep{Liu2019,Casamiquela2020}.  A novel method of bypassing systematic uncertainties devised by \citet{Bovy2016a} expands the line-by-line differential analysis beyond solar twins.  In this method, stellar spectra are parametrized as functions of $\rm T_{eff}$ before being analyzed differentially for spectrum-to-spectrum differences \citep[e.g.,][for more details]{Cheng2020, deMijolla2021}.

Similar to the methods employed by studies with comparable sample sizes, we choose to employ MLE \citep[][]{Walker2006, Piatti2018, Ness2018, Kovalev2019, Poovelil2020}.  In the context of our study, MLE enables us to determine the most likely parent distribution from which the observed abundances were drawn.  In this case, we assume that the parent [X/H] distributions that describe the chemical profile of each structure are Gaussian in nature with mean abundance $\rm \mu_{[X/H]}$ and intrinsic dispersion $\rm \sigma_{[X/H]}$.  We aim to find the values of $\rm \mu_{[X/H]}, \sigma_{[X/H]}$ that maximize the following likelihood function:

\begin{equation}
    \mathcal{L} = \prod_i^N \rm{exp} \left [ \frac{-(x_i - \mu_{[X/H]})^2} {(2(\sigma_{[X/H]}^2 + \delta_i^2)} \right ] \times \frac{1}{\sqrt{2\pi(\sigma_{[X/H]}^2 + \delta_i^2)}}
\end{equation}

where N is the number of samples per cluster, $\rm x_i$ is the measured abundance for each star, and $\delta_i$ is the GALAH abundance uncertainty.  We validate that this MLE retrieves accurate intrinsic mean and dispersion measurements of artificial abundance data.

We report our MLE-derived intrinsic dispersions in Table \ref{tab:intrinsic}.  Uncertainties on the intrinsic dispersion are reported as the 16th and 84th percentiles of the intrinsic dispersion likelihood function.  We also derive the intrinsic dispersions of both the local field surrounding each structure and M 67 using the same MLE.

In order for our MLE to report accurate intrinsic dispersion estimates, it is crucial that the GALAH abundance uncertainties input into the MLE are not over- or underestimated.  Over- or underestimating abundance uncertainties results in under- or overestimated intrinsic dispersions, respectively.  Accuracy in our MLE intrinsic dispersion results relies on the assumption that the GALAH abundance uncertainties are appropriate and not over- or underestimates. To test the robustness of our MLE results to errors in reported uncertainty, we re-compute our results with a hypothetical 50\% over- or underestimation of the GALAH abundance uncertainties.  We find that this degree of over- or underestimation of the uncertainties would still result in intrinsic dispersions that lie within the $\rm 1\sigma$ uncertainties of our original MLE results in Table \ref{tab:intrinsic}.  We refer readers to \citet{GALAHDR3}, Section 4, for a validation of the reported GALAH abundance uncertainties and a description of their robust method for determining them.

\subsection{Using Chemistry to Probe Youth and Age}\label{subsec:lithium}
In addition to allowing us insight into the chemical homogeneity of each structure, the surface abundances reported by GALAH DR3 enable us to validate isochronal ages derived by KC19.  Li is one such element that allows us to probe the youth of the structures in our sample.  The surface abundance of Li, denoted here as $\rm A(Li) = log_{10}\left(\frac{N_{Li}}{N_H}\right) + 12.00$, is an indicator of youth in low- and intermediate-mass stars \citep{Dantona1984, Soderblom1993, Baumann2010, Monroe2013, Melendez2014, TucciMaia2015}. It is destroyed at the relatively low temperatures ($\sim$ 2.5 $\times$ 10$^6$K) found at the base of the convective envelopes of low- and intermediate-mass stars \citep[e.g.,][]{Soderblom2014, Gutierrez2020}.  As such, stars with convective envelopes deplete their surface Li over time.  The rate of Li depletion is highly dependent on the thickness of the convective envelope, where stars with thicker envelopes deplete Li much more rapidly.  Since the mass of the convective envelope correlates directly with $\rm T_{eff}$, at a fixed age, cooler stars display a greater depletion in Li relative to hotter stars.  Thus, one can construct age-dependent Li tracks in the A(Li)-$\rm T_{eff}$ plane to estimate stellar ages \citep{Soderblom2014}.  These tracks can either be constructed theoretically using stellar evolution models \citep[e.g.,][]{Charbonnel2005, Andrassy2015, Carlos2016,  Dumont2020} or empirically using the Li abundances of stars in coeval populations \citep[e.g.,][]{Hobbs1986, Hobbs1988, Hawkins2020, Gutierrez2020}.  We note that though surface Li is depleted with age due to convection in a pattern that tends to scale with $\rm T_{eff}$, it is also sensitive to stellar rotation, metallicity, and planet occurrence, all of which can alter the shape of the Li track \citep[e.g.,][]{Montalban2002, Sandquist2002}. Thus, we use these Li tracks as tools to generally estimate youth rather than to constrain ages to high precisions.

To interpret the GALAH Li data of each KC19 structure, we construct empirical Li tracks in the A(Li)-$\rm T_{eff}$ plane that act to anchor the Li distributions of each structure in age space.  We construct these tracks by fitting univariate splines (using the \code{scipy.interpolate} package with a smoothing factor of 4.5) to the A(Li) vs. $\rm T_{eff}$ distributions of several GALAH-sampled OCs with well-constrained ages.  To obtain GALAH data for several OCs, we crossmatch the OC membership catalog of \citet{Spina2021} with GALAH DR3 using the quality cuts defined in Section \ref{subsec:GALAHDR3}.  We ultimately create empirical Li tracks from six OCs that have at least six high-quality GALAH-sampled stars with T$\rm _{eff}$ \textless 6500 K: ASCC 16 \citep[13.0$\pm$1.3 Myr,][]{Kos2019}, IC 4665 \citep[36$\pm$9 Myr,][]{Miret-Roig2019}, Blanco 1 \citep[ $100^{+46}_{-20}$ Myr,][]{Zhang2020}, Pleiades \citep[100$^{+41}_{-23}$ Myr,][]{Bell2012, Scholz2015, Dahm2015, Yen2018, Gossage2018, Bossini2019}, Hyades \citep[625$\pm$50 Myr, e.g.,][]{Perryman1998, Gossage2018, GaiaDR2}, and M 67 \citep[3.75$^{+1.25}_{-0.50}$ Gyr, e.g.,][]{Stello2016, Gao2018, Sandquist2021}.  The membership data for M 67 is that of \citet{Gao2018} to maintain consistency in our consideration of this particular OC. We compare the distribution of A(Li) vs. $\rm T_{eff}$ for each KC19 cluster with the empirical calibration A(Li) vs. $\rm T_{eff}$ tracks to validate the isochronally-derived ages of KC19.

In addition to validating structure youth with Li, we also investigate the age of each cluster using \textit{chemical clocks}.  Chemical clocks are elemental abundance ratios that have been observed to correlate with stellar age \citep[e.g.,][]{Nissen2015, Jofre2020, Hayden2020, Espinoza-Rojas2021}.  By comparing the abundance ratios of two elements with two different nucleosynthetic timescales, such as $\rm \alpha$-elements, which are produced in the core-collapse supernovae of short-lived high mass stars, and s-process elements, which are mainly produced in long-lived, low- and intermediate-mass AGB stars, one can probe the relative age of a star.  Chemical clocks have been shown to scale with age in the Solar neighborhood (<1 kpc of the Sun) for stars with Solar metallicity \citep[e.g.,][]{Nissen2015, Jofre2020, Espinoza-Rojas2021}.  Thus, we use chemical clocks as an additional tool to probe the relative ages of the clusters investigated here.

To select the chemical clocks for this portion of our study, we refer to the results of \citet{Espinoza-Rojas2021}.  This work studied the consistency of chemical clocks in wide binaries across a broad stellar parameter space and found that several specific chemical clocks are extremely consistent between coeval stars, even when [X/Fe] abundances are not.  We examine three of the most consistent chemical clocks from their study within our sample, namely [Sc/Ba], [Ca/Ba], [Ti/Ba].  In addition, we include [Mg/Y] in our analysis, a well-studied chemical clock that has been found to strongly correlate with age in Solar-type stars \citep[e.g.,][]{Nissen2015, Spina2016, Tucci2016, Bedell2018, Spina2018}.  We qualitatively investigate the correlation between isochronal age and chemical clock abundance in our sample to further validate the isochronal ages of each structure.

\section{Results}\label{Sec:Results}

\begin{figure*}
    \centering
    \includegraphics[width=2\columnwidth]{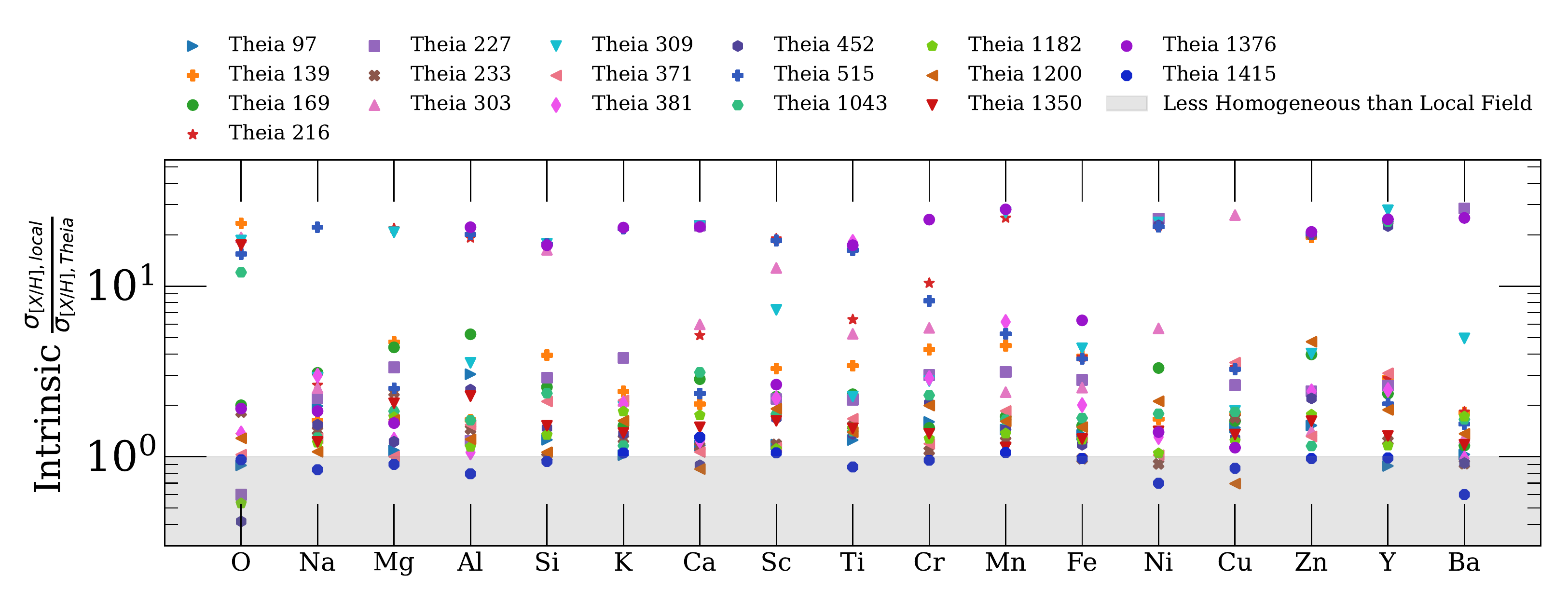}
    \caption{The ratio of intrinsic chemical dispersion of the local field surrounding each KC19 structure relative to that of each structure (y-axis) across 17 elements (x-axis).  The average uncertainty in this ratio is $\pm$0.08.  The gray shading demarcates the region of the plot where the intrinsic dispersion of a structure equals or surpasses that of its local field.  Larger y-axis values indicate that the structure is \textit{more} homogeneous than its local field (white, unshaded region).  Most structures are more homogeneous in most elements than their local fields.  The one major exception to this is Theia 1415, which is generally less homogeneous than its local field in all elements but Ca.}\label{fig:violin1}
\end{figure*}

\begin{figure*}
	\includegraphics[width=2\columnwidth]{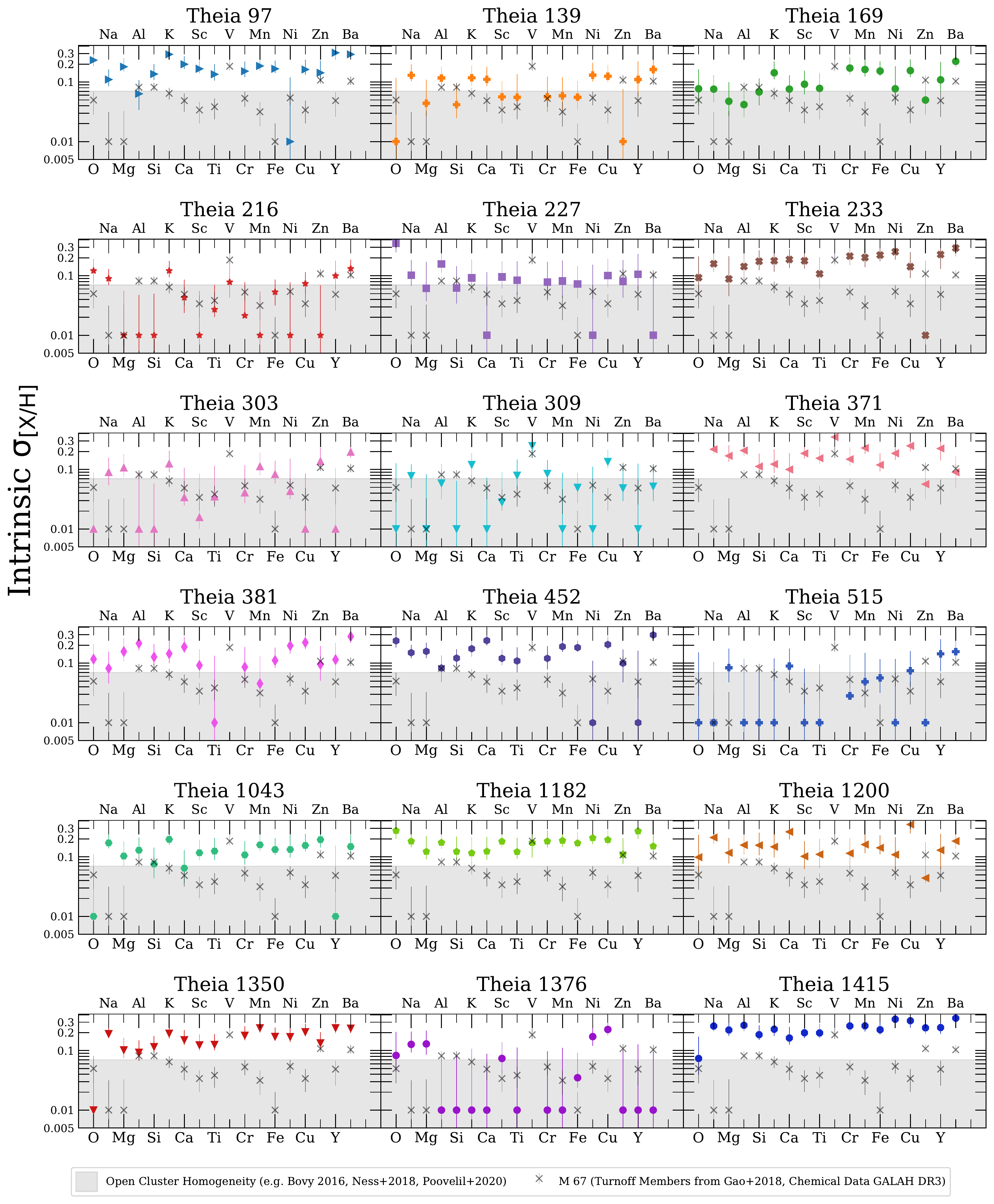}
     \caption{The intrinsic chemical dispersions in various elements for each KC19 group (colored, solid markers).  The gray shaded region demarcates the region of reported OC dispersions \citep[e.g.,][]{Bovy2016a, Ness2018, Poovelil2020}.  We compare the intrinsic dispersions in [X/H] of each structure to those of M 67, a well-studied OC \citep[e.g.,][]{Bovy2016a, Casamiquela2020}, shown with gray x's.  Several structures, including Theias 139, 169, 216, 227, 303, 309, 515, and 1376, possess intrinsic dispersions in [X/H] similar to those reported in OCs in several elements.}\label{fig:OCviolin2}
 \end{figure*}

{\renewcommand{\arraystretch}{1.5}
\begin{table*}
\begin{tabular}{c|ccccccccc||c||}
\hline
Theia & 97 & 139 & 169 & 216 & 227 & 233 & 303 & 309 & 371 & M 67\\
\hline
$\rm \sigma_{[O/H]}$ & $0.23^{+0.07}_{-0.05}$ & $0.01^{+0.11}_{-0.01}$ & $0.08^{+0.09}_{-0.04}$ & $0.12^{+0.07}_{-0.05}$ & $0.35^{+0.02}_{-0.10}$ & $0.09^{+0.12}_{-0.04}$ & $0.01^{+0.08}_{-0.01}$ & $0.01^{+0.12}_{-0.02}$ &  & $0.05^{+0.02}_{-0.02}$ \\
$\rm \sigma_{[Na/H]}$ & $0.11^{+0.05}_{-0.03}$ & $0.13^{+0.07}_{-0.03}$ & $0.08^{+0.06}_{-0.03}$ & $0.09^{+0.04}_{-0.02}$ & $0.10^{+0.08}_{-0.03}$ & $0.16^{+0.10}_{-0.04}$ & $0.09^{+0.07}_{-0.04}$ & $0.08^{+0.04}_{-0.03}$ & $0.22^{+0.08}_{-0.04}$ & $0.01^{+0.02}_{-0.01}$\\
$\rm \sigma_{[Mg/H]}$ & $0.18^{+0.07}_{-0.03}$ & $0.04^{+0.07}_{-0.02}$ & $0.05^{+0.05}_{-0.02}$ & $0.01^{+0.05}_{-0.01}$ & $0.06^{+0.11}_{-0.02}$ & $0.09^{+0.12}_{-0.04}$ & $0.11^{+0.07}_{-0.03}$ & $0.01^{+0.07}_{-0.01}$ & $0.17^{+0.09}_{-0.03}$ & $0.01^{+0.02}_{-0.01}$\\
$\rm \sigma_{[Al/H]}$ & $0.06^{+0.04}_{-0.03}$ & $0.12^{+0.07}_{-0.02}$ & $0.04^{+0.06}_{-0.02}$ & $0.01^{+0.04}_{-0.0}$ & $0.16^{+0.10}_{-0.05}$ & $0.14^{+0.10}_{-0.05}$ & $0.01^{+0.08}_{-0.01}$ & $0.06^{+0.06}_{-0.03}$ & $0.21^{+0.08}_{-0.04}$ & $0.08^{+0.02}_{-0.01}$\\
$\rm \sigma_{[Si/H]}$ & $0.14^{+0.06}_{-0.03}$ & $0.04^{+0.05}_{-0.02}$ & $0.07^{+0.05}_{-0.03}$ & $0.01^{+0.04}_{-0.0}$ & $0.06^{+0.08}_{-0.03}$ & $0.17^{+0.09}_{-0.05}$ & $0.01^{+0.05}_{-0.01}$ & $0.01^{+0.05}_{-0.01}$ & $0.11^{+0.07}_{-0.03}$ & $0.08^{+0.02}_{-0.01}$\\
$\rm \sigma_{[K/H]}$ & $0.29^{+0.03}_{-0.05}$ & $0.12^{+0.07}_{-0.03}$ & $0.14^{+0.08}_{-0.03}$ & $0.12^{+0.06}_{-0.03}$ & $0.09^{+0.10}_{-0.04}$ & $0.18^{+0.09}_{-0.06}$ & $0.12^{+0.08}_{-0.04}$ & $0.12^{+0.07}_{-0.05}$ & $0.12^{+0.10}_{-0.04}$ & $0.06^{+0.02}_{-0.01}$\\
$\rm \sigma_{[Ca/H]}$ & $0.20^{+0.07}_{-0.03}$ & $0.11^{+0.07}_{-0.03}$ & $0.08^{+0.05}_{-0.03}$ & $0.04^{+0.04}_{-0.02}$ & $0.01^{+0.10}_{-0.01}$ & $0.19^{+0.09}_{-0.05}$ & $0.03^{+0.07}_{-0.01}$ & $0.01^{+0.06}_{-0.01}$ & $0.10^{+0.09}_{-0.05}$ & $0.05^{+0.02}_{-0.02}$\\
$\rm \sigma_{[Sc/H]}$ & $0.17^{+0.06}_{-0.03}$ & $0.06^{+0.05}_{-0.02}$ & $0.09^{+0.06}_{-0.03}$ & $0.01^{+0.05}_{-0.01}$ & $0.10^{+0.08}_{-0.03}$ & $0.18^{+0.09}_{-0.05}$ & $0.02^{+0.07}_{-0.01}$ & $0.03^{+0.06}_{-0.01}$ & $0.19^{+0.08}_{-0.03}$ & $0.03^{+0.01}_{-0.01}$\\
$\rm \sigma_{[Ti/H]}$ & $0.14^{+0.07}_{-0.03}$ & $0.06^{+0.08}_{-0.02}$ & $0.08^{+0.06}_{-0.03}$ & $0.03^{+0.04}_{-0.01}$ & $0.08^{+0.09}_{-0.04}$ & $0.11^{+0.10}_{-0.05}$ & $0.04^{+0.08}_{-0.01}$ & $0.08^{+0.07}_{-0.04}$ & $0.15^{+0.08}_{-0.03}$ & $0.04^{+0.01}_{-0.01}$\\
$\rm \sigma_{[Cr/H]}$ & $0.15^{+0.07}_{-0.03}$ & $0.06^{+0.07}_{-0.02}$ & $0.17^{+0.09}_{-0.04}$ & $0.02^{+0.06}_{-0.01}$ & $0.08^{+0.09}_{-0.04}$ & $0.21^{+0.08}_{-0.06}$ & $0.04^{+0.08}_{-0.01}$ & $0.09^{+0.06}_{-0.04}$ & $0.15^{+0.09}_{-0.04}$ & $0.05^{+0.02}_{-0.01}$\\
$\rm \sigma_{[Mn/H]}$ & $0.19^{+0.07}_{-0.03}$ & $0.06^{+0.06}_{-0.03}$ & $0.16^{+0.09}_{-0.04}$ & $0.01^{+0.04}_{-0.01}$ & $0.08^{+0.10}_{-0.04}$ & $0.20^{+0.09}_{-0.06}$ & $0.11^{+0.08}_{-0.03}$ & $0.01^{+0.08}_{-0.01}$ & $0.23^{+0.07}_{-0.05}$ & $0.03^{+0.01}_{-0.01}$\\
$\rm \sigma_{[Fe/H]}$ & $0.17^{+0.06}_{-0.02}$ & $0.06^{+0.05}_{-0.02}$ & $0.15^{+0.07}_{-0.03}$ & $0.05^{+0.03}_{-0.02}$ & $0.07^{+0.07}_{-0.03}$ & $0.22^{+0.08}_{-0.05}$ & $0.08^{+0.06}_{-0.03}$ & $0.05^{+0.04}_{-0.02}$ & $0.12^{+0.07}_{-0.03}$ & $0.01^{+0.01}_{-0.01}$\\
$\rm \sigma_{[Ni/H]}$ & $0.01^{+0.11}_{-0.01}$ & $0.13^{+0.08}_{-0.03}$ & $0.08^{+0.11}_{-0.03}$ & $0.01^{+0.07}_{-0.01}$ & $0.01^{+0.14}_{-0.02}$ & $0.25^{+0.06}_{-0.07}$ & $0.04^{+0.11}_{-0.01}$ & $0.01^{+0.10}_{-0.01}$ & $0.19^{+0.09}_{-0.04}$ & $0.05^{+0.01}_{-0.01}$\\
$\rm \sigma_{[Cu/H]}$ & $0.16^{+0.07}_{-0.03}$ & $0.12^{+0.07}_{-0.02}$ & $0.16^{+0.08}_{-0.03}$ & $0.07^{+0.04}_{-0.03}$ & $0.10^{+0.09}_{-0.04}$ & $0.14^{+0.11}_{-0.05}$ & $0.01^{+0.08}_{-0.01}$ & $0.14^{+0.08}_{-0.03}$ & $0.25^{+0.06}_{-0.05}$ & $0.03^{+0.01}_{-0.01}$\\
$\rm \sigma_{[Zn/H]}$ & $0.14^{+0.09}_{-0.04}$ & $0.01^{+0.07}_{-0.01}$ & $0.05^{+0.06}_{-0.02}$ & $0.01^{+0.06}_{-0.01}$ & $0.08^{+0.10}_{-0.04}$ & $0.01^{+0.17}_{-0.02}$ & $0.14^{+0.09}_{-0.04}$ & $0.05^{+0.07}_{-0.02}$ & $0.06^{+0.08}_{-0.02}$ & $0.11^{+0.02}_{-0.01}$\\
$\rm \sigma_{[Y/H]}$ & $0.31^{+0.02}_{-0.07}$ & $0.11^{+0.11}_{-0.05}$ & $0.11^{+0.11}_{-0.04}$ & $0.10^{+0.08}_{-0.04}$ & $0.11^{+0.13}_{-0.05}$ & $0.23^{+0.07}_{-0.10}$ & $0.01^{+0.10}_{-0.01}$ & $0.01^{+0.12}_{-0.01}$ & $0.22^{+0.08}_{-0.08}$ & $0.05^{+0.02}_{-0.02}$\\
$\rm \sigma_{[Ba/H]}$ & $0.29^{+0.03}_{-0.05}$ & $0.16^{+0.08}_{-0.04}$ & $0.22^{+0.07}_{-0.04}$ & $0.13^{+0.06}_{-0.02}$ & $0.01^{+0.10}_{-0.01}$ & $0.29^{+0.03}_{-0.08}$ & $0.20^{+0.08}_{-0.04}$ & $0.05^{+0.06}_{-0.02}$ & $0.09^{+0.08}_{-0.04}$ & $0.10^{+0.02}_{-0.01}$\\

\hline
\hline
Theia & 381 & 452 & 515 & 1043 & 1182 & 1200 & 1350 & 1376 & \multicolumn{2}{l|}{1415} \\
 \hline
$\rm \sigma_{[O/H]}$ & $0.12^{+0.09}_{-0.05}$ & $0.24^{+0.06}_{-0.06}$ & $0.01^{+0.14}_{-0.02}$ & $0.01^{+0.10}_{-0.01}$ & $0.27^{+0.04}_{-0.07}$ & $0.10^{+0.14}_{-0.05}$ & $0.01^{+0.07}_{-0.01}$ & $0.08^{+0.12}_{-0.04}$ & \multicolumn{2}{l|}{$0.07^{+0.10}_{-0.03}$} \\
$\rm \sigma_{[Na/H]}$ & $0.08^{+0.08}_{-0.04}$ & $0.15^{+0.05}_{-0.03}$ & $0.01^{+0.09}_{-0.01}$ & $0.17^{+0.08}_{-0.03}$ & $0.18^{+0.09}_{-0.04}$ & $0.21^{+0.08}_{-0.05}$ & $0.19^{+0.08}_{-0.03}$ & $0.13^{+0.08}_{-0.03}$ & \multicolumn{2}{l|}{$0.26^{+0.06}_{-0.05}$} \\
$\rm \sigma_{[Mg/H]}$ & $0.16^{+0.10}_{-0.05}$ & $0.16^{+0.06}_{-0.04}$ & $0.08^{+0.09}_{-0.04}$ & $0.10^{+0.08}_{-0.03}$ & $0.12^{+0.10}_{-0.03}$ & $0.12^{+0.09}_{-0.04}$ & $0.10^{+0.06}_{-0.03}$ & $0.13^{+0.09}_{-0.04}$ & \multicolumn{2}{l|}{$0.22^{+0.07}_{-0.04}$} \\
$\rm \sigma_{[Al/H]}$ & $0.21^{+0.08}_{-0.05}$ & $0.08^{+0.05}_{-0.04}$ & $0.01^{+0.11}_{-0.01}$ & $0.13^{+0.08}_{-0.03}$ & $0.17^{+0.10}_{-0.04}$ & $0.16^{+0.10}_{-0.05}$ & $0.09^{+0.06}_{-0.03}$ & $0.01^{+0.07}_{-0.01}$ & \multicolumn{2}{l|}{$0.27^{+0.05}_{-0.05}$} \\
$\rm \sigma_{[Si/H]}$ & $0.13^{+0.08}_{-0.03}$ & $0.12^{+0.05}_{-0.03}$ & $0.01^{+0.07}_{-0.01}$ & $0.08^{+0.07}_{-0.03}$ & $0.12^{+0.10}_{-0.03}$ & $0.16^{+0.10}_{-0.05}$ & $0.12^{+0.06}_{-0.02}$ & $0.01^{+0.07}_{-0.01}$ & \multicolumn{2}{l|}{$0.19^{+0.08}_{-0.03}$} \\
$\rm \sigma_{[K/H]}$ & $0.14^{+0.09}_{-0.04}$ & $0.17^{+0.07}_{-0.05}$ & $0.01^{+0.11}_{-0.01}$ & $0.20^{+0.08}_{-0.04}$ & $0.12^{+0.10}_{-0.04}$ & $0.15^{+0.09}_{-0.05}$ & $0.19^{+0.08}_{-0.03}$ & $0.01^{+0.10}_{-0.01}$ & \multicolumn{2}{l|}{$0.23^{+0.07}_{-0.05}$} \\
$\rm \sigma_{[Ca/H]}$ & $0.19^{+0.09}_{-0.04}$ & $0.24^{+0.06}_{-0.04}$ & $0.09^{+0.09}_{-0.04}$ & $0.06^{+0.07}_{-0.03}$ & $0.12^{+0.09}_{-0.03}$ & $0.26^{+0.05}_{-0.07}$ & $0.15^{+0.07}_{-0.03}$ & $0.01^{+0.08}_{-0.01}$ & \multicolumn{2}{l|}{$0.16^{+0.09}_{-0.04}$} \\
$\rm \sigma_{[Sc/H]}$ & $0.09^{+0.08}_{-0.04}$ & $0.12^{+0.05}_{-0.03}$ & $0.01^{+0.07}_{-0.01}$ & $0.12^{+0.08}_{-0.03}$ & $0.18^{+0.09}_{-0.04}$ & $0.10^{+0.08}_{-0.04}$ & $0.12^{+0.06}_{-0.02}$ & $0.07^{+0.06}_{-0.03}$ & \multicolumn{2}{l|}{$0.20^{+0.08}_{-0.04}$} \\
$\rm \sigma_{[Ti/H]}$ & $0.01^{+0.12}_{-0.02}$ & $0.11^{+0.08}_{-0.04}$ & $0.01^{+0.09}_{-0.01}$ & $0.12^{+0.08}_{-0.04}$ & $0.12^{+0.09}_{-0.03}$ & $0.11^{+0.09}_{-0.05}$ & $0.13^{+0.06}_{-0.02}$ & $0.01^{+0.10}_{-0.01}$ & \multicolumn{2}{l|}{$0.20^{+0.09}_{-0.04}$} \\
$\rm \sigma_{[Cr/H]}$ & $0.09^{+0.10}_{-0.04}$ & $0.12^{+0.07}_{-0.04}$ & $0.03^{+0.11}_{-0.01}$ & $0.11^{+0.08}_{-0.03}$ & $0.18^{+0.09}_{-0.04}$ & $0.11^{+0.09}_{-0.05}$ & $0.18^{+0.08}_{-0.03}$ & $0.01^{+0.09}_{-0.01}$ & \multicolumn{2}{l|}{$0.26^{+0.06}_{-0.05}$} \\
$\rm \sigma_{[Mn/H]}$ & $0.05^{+0.09}_{-0.02}$ & $0.19^{+0.07}_{-0.04}$ & $0.05^{+0.10}_{-0.02}$ & $0.16^{+0.09}_{-0.04}$ & $0.19^{+0.09}_{-0.04}$ & $0.16^{+0.09}_{-0.05}$ & $0.24^{+0.07}_{-0.04}$ & $0.01^{+0.10}_{-0.01}$ & \multicolumn{2}{l|}{$0.26^{+0.05}_{-0.05}$} \\
$\rm \sigma_{[Fe/H]}$ & $0.11^{+0.07}_{-0.03}$ & $0.18^{+0.06}_{-0.03}$ & $0.06^{+0.06}_{-0.02}$ & $0.13^{+0.07}_{-0.03}$ & $0.17^{+0.09}_{-0.03}$ & $0.14^{+0.09}_{-0.03}$ & $0.17^{+0.07}_{-0.03}$ & $0.04^{+0.06}_{-0.01}$ & \multicolumn{2}{l|}{$0.22^{+0.07}_{-0.04}$} \\
$\rm \sigma_{[Ni/H]}$ & $0.20^{+0.09}_{-0.05}$ & $0.01^{+0.10}_{-0.01}$ & $0.01^{+0.11}_{-0.01}$ & $0.13^{+0.10}_{-0.04}$ & $0.21^{+0.08}_{-0.04}$ & $0.11^{+0.10}_{-0.05}$ & $0.17^{+0.08}_{-0.03}$ & $0.17^{+0.09}_{-0.05}$ & \multicolumn{2}{l|}{$0.34^{+0.01}_{-0.10}$} \\
$\rm \sigma_{[Cu/H]}$ & $0.22^{+0.08}_{-0.05}$ & $0.21^{+0.07}_{-0.04}$ & $0.07^{+0.08}_{-0.04}$ & $0.16^{+0.09}_{-0.03}$ & $0.19^{+0.09}_{-0.04}$ & $0.35^{+0.02}_{-0.10}$ & $0.21^{+0.08}_{-0.03}$ & $0.23^{+0.07}_{-0.05}$ & \multicolumn{2}{l|}{$0.32^{+0.01}_{-0.08}$} \\
$\rm \sigma_{[Zn/H]}$ & $0.10^{+0.10}_{-0.04}$ & $0.10^{+0.07}_{-0.05}$ & $0.01^{+0.11}_{-0.01}$ & $0.19^{+0.09}_{-0.05}$ & $0.11^{+0.10}_{-0.03}$ & $0.04^{+0.13}_{-0.01}$ & $0.13^{+0.07}_{-0.03}$ & $0.01^{+0.09}_{-0.01}$ & \multicolumn{2}{l|}{$0.24^{+0.07}_{-0.05}$} \\
$\rm \sigma_{[Y/H]}$ & $0.11^{+0.10}_{-0.05}$ & $0.01^{+0.15}_{-0.02}$ & $0.14^{+0.11}_{-0.07}$ & $0.01^{+0.14}_{-0.02}$ & $0.27^{+0.05}_{-0.07}$ & $0.13^{+0.11}_{-0.07}$ & $0.24^{+0.07}_{-0.05}$ & $0.01^{+0.12}_{-0.01}$ & \multicolumn{2}{l|}{$0.24^{+0.06}_{-0.05}$} \\
$\rm \sigma_{[Ba/H]}$ & $0.28^{+0.04}_{-0.06}$ & $0.30^{+0.03}_{-0.05}$ & $0.16^{+0.10}_{-0.06}$ & $0.15^{+0.09}_{-0.04}$ & $0.15^{+0.10}_{-0.04}$ & $0.18^{+0.09}_{-0.05}$ & $0.24^{+0.07}_{-0.04}$ & $0.01^{+0.08}_{-0.01}$ & \multicolumn{2}{l|}{$0.35^{+0.01}_{-0.07}$} \\

\hline
 \hline
\end{tabular}
\caption{Intrinsic dispersions in [X/H] derived using Maximum Likelihood Estimation for each structure and for M 67, the reference OC.  We only consider elements with at least 6 high-quality (see Section \ref{subsec:GALAHDR3}) abundance measurements reported by GALAH.  Errors on the intrinsic dispersions are presented as the dispersion values associated with the 16th and 84th percentiles of the resulting likelihood function (see Section \ref{subsec:defininghomogeneity}), displayed as sub- and superscripts, respectively.  We omit $\rm \sigma_{[O/H]}$ for Theia 371 because it has fewer than 6 stars with high-quality O abundances reported by GALAH.} \label{tab:intrinsic}
\end{table*}
}

\begin{figure}
	\includegraphics[width=1\columnwidth]{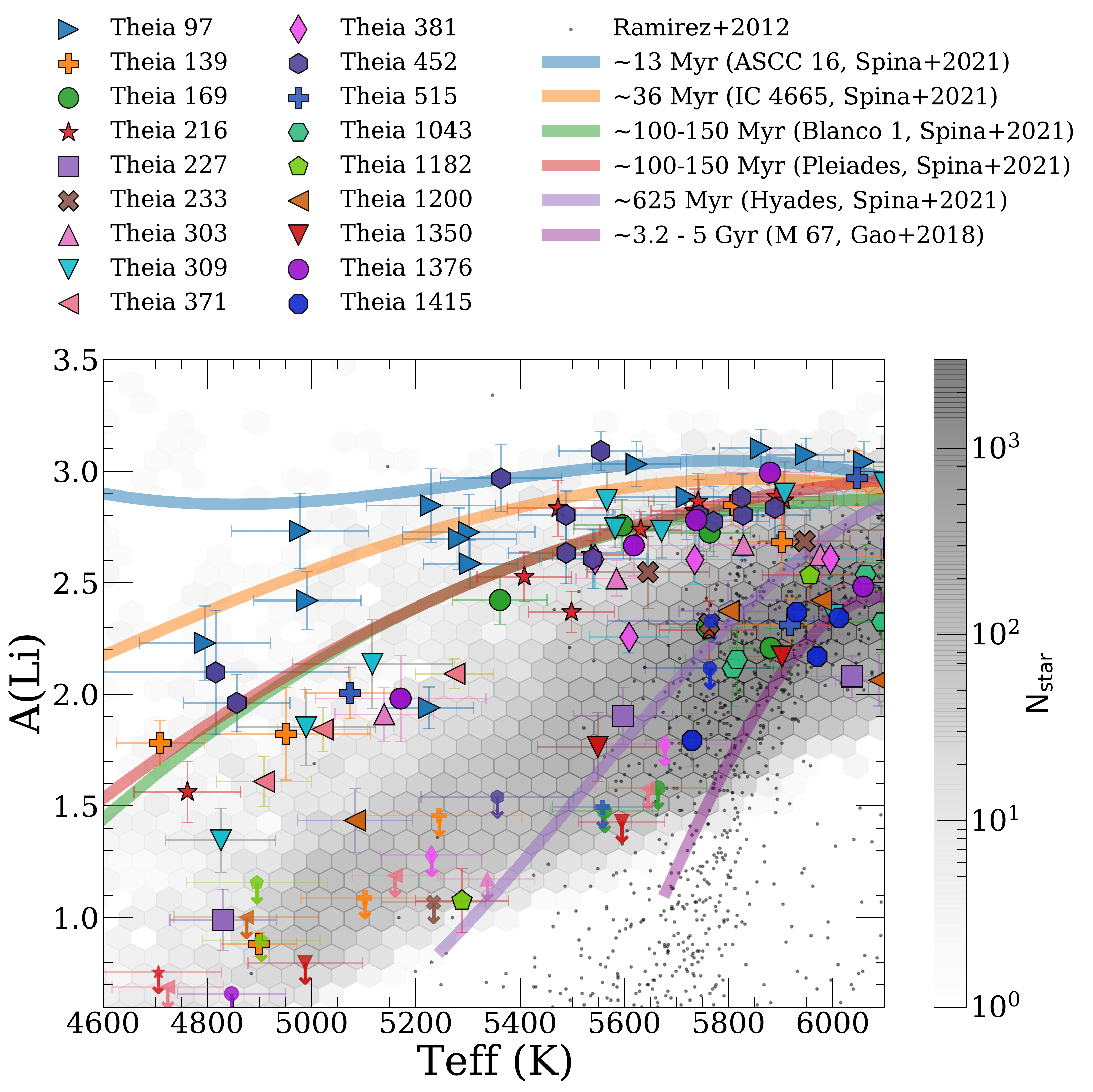}
    \caption{A(Li) vs. $\rm T_{eff}$ for all high-quality GALAH-sampled KC19 structures investigated in this study (large symbols).  Smaller symbols with downward arrows indicate that the Li abundance is reported as an upper-limit.  The solid lines represent empirically-derived A(Li) vs. $\rm T_{eff}$ tracks using GALAH-reported Li abundances of various OCs with ages reported in the literature and memberships adopted from \citet{Gao2018} (M 67)  and \citet{Spina2021} (remaining OCs).  The background density distribution represents the GALAH-sampled local field, which we define as all FGK dwarfs within 1 kpc of the Sun that do not belong to a known OC or association.  We also include non-GALAH data from \citet{Ramirez2012}, which reports the A(Li) vs. $\rm T_{eff}$ distribution of disk FGK dwarfs within 200 pc of the Sun for comparison.  Theias 97 and 452 are among the youngest of our sample, while Theia 1415 is the oldest.  Most of the structures we study here have Li depletion patterns that largely agree with their isochronal ages with the exception of Theia 227, which appears chemically older, and Theias 452 and 1376, which appear chemically younger.}
    \label{fig:liteff}
\end{figure}

\begin{table}
	\centering
	\caption{KC19 isochronal ages (age uncertainties of $\sim$0.15 dex) and Li depletion ages of each structure.  Generally, the Li abundances of each structure agree with the isochronal ages. }
	\label{tab:liages}
	\begin{tabular}{lcr} 
		\hline
		Theia & Isochronal Age$^{a}$ & Lithium Depletion Age \\
		    & (Myr) & (Myr)  \\
		\hline
		97 & 33 & 21 -- 100 \\
        139  & 51 &  100--625 \\
        169  & 79 &  100--150  \\
        216  & 97 &  100--150 \\
        227  & 97 & 150--625 \\
        233  & 80  &  150--625 \\
        303  & 108 & $\sim$150  \\
        309  & 109 & $\sim$150  \\
        371  & 146 & $\sim$150  \\
        381  & 130 & 150--625  \\
        452  & 171 &   21--150 \\
        515  & 200 & 150--625  \\
        1043  & 767 &  >625 \\
        1182  & 1,000 &  $\geq$625 \\
        1200  & 1,159 &  $\geq$625 \\
        1350  & 1,600 &   \textgreater625 \\
        1376  & 1,600 &  150--625  \\
        1415  & 1,720 &  $\geq$625  \\
		\hline
	\end{tabular}
	
$^{a}$ Taken from Table~2 of KC19 
\end{table}

A major component of our analysis involves comparing the intrinsic dispersions in [X/H] of each structure to those of their local fields and the well-studied OC M 67 \citep[e.g.,][]{Bovy2016a, Poovelil2020}.  We estimate intrinsic dispersions using the MLE described in Section \ref{subsec:defininghomogeneity}, which disentangles the intrinsic abundance distribution from the observed distribution to probe the inherent [X/H] dispersions of each cluster.  We present the results of our MLE in Table \ref{tab:intrinsic}, which tabulates our intrinsic dispersion estimates in up to 17 elements for each structure.  These results can be compared to the homogeneity of both other stellar populations reported in the literature \citep[e.g.,][]{DeSilva2007, Barenfeld2013, Bovy2016a, Lambert2016, Ness2018, Kovalev2019, Poovelil2020} and the MW ISM \citep[e.g.,][]{Cartledge2006, Przybilla2008, Cordova2020} to place the intrinsic dispersions in [X/H] of the KC19 structures in context.  The following subsections present our general results in the context of our full sample, but we refer readers to Appendix \ref{appendixa} for more detailed descriptions of our findings on a structure-by-structure basis.

\subsection{Comparison of Intrinsic Dispersions of Structures to Those of Local Field and M 67}
We begin by comparing the intrinsic dispersions in [X/H] of each structure with those of their local fields (see Section \ref{subsec:localfield} for the method we use to define each local field).  Figure \ref{fig:violin1} presents the ratio of the intrinsic dispersions in [X/H] of the local field of each structure to those of each structure.  Structures with dispersion ratios less than or equal to 1 (gray shaded region) possess an intrinsic dispersion that equals or exceeds that of their field, suggesting that they are chemically indistinguishable from their local stellar neighborhood.  Structures with ratios greater than 1 are more chemically homogeneous than their local field.  The average uncertainty in this ratio is $\pm$ 0.08, which we determine by adding in quadrature the MLE-reported uncertainties in the field and structure dispersions, and it is dominated by the uncertainties in the intrinsic dispersions for each structure, as the uncertainties in the intrinsic dispersions of the local fields are extremely small (\textless 0.007 dex) due to the large population sizes (on average, $\rm 4200 \pm 2400~stars$ per local field cylinder).  We find that almost all structures in our sample have intrinsic dispersions in [X/H] that are lower than those of their local field by a factor of 1.5 to 30 times, which suggests that most structures are more homogeneous than their local fields.  There are some exceptions to this, however, with structures possessing intrinsic dispersions in certain elements that exceed those of its local field.  The most extreme exception is Theia 1415, which is one of the least homogeneous structures in our sample ($\rm \sigma_{[X/H]}$ between 0.16 and 0.35 dex in most elements) and exceeds the intrinsic dispersions in [X/H] of its field in all elements but Ca.

To further contextualize the intrinsic dispersions in [X/H] of each KC19 cluster, we also compare them with those of M 67.  In this way, we can investigate how the intrinsic dispersions in chemical abundance compare with those of a well-studied OC observed by the same survey.  Figure \ref{fig:OCviolin2} presents the comparison of intrinsic dispersions in [X/H] of each structure (colored symbols) with those of M 67 (denoted with x's).  We also shade in gray the range of intrinsic dispersions in [X/H] observed in OCs in other works \citep[e.g.,][]{Lambert2016, Ness2018, Kovalev2019, Casamiquela2020, Poovelil2020}.  
We find a range of intrinsic dispersions in [X/H] among the structures.  Five strings, Theias 139, 169, 216, 303, and 309, are as chemically homogeneous in the majority of the elements as OCs.  The remaining five strings (Theias 97, 233, 452, 1200, and 1415) are less homogeneous than OCs, with intrinsic dispersions in [X/H] that range from 0.08 dex to 0.35 dex.  We also find a wide range of intrinsic dispersions in [X/H] in the compact, non-string-like structures in our sample.  Theias 227, 515, and 1376 each show intrinsic dispersions that are comparable to those found in OCs in most elements.  By contrast, the remaining five compact structures in our sample (Theias 371, 381, 1043, 1182, and 1350) show high intrinsic dispersions in [X/H] (between 0.1 and 0.35 dex) that exceed those found in OCs.  We present these results in greater detail on a structure-by-structure basis in Appendix \ref{appendixa}.

\subsection{No Apparent Correlations between Intrinsic Dispersion in [X/H] and Length}\label{subsec:badisp}
In seeking a possible explanation for these variations in intrinsic dispersions among the structures, we search for a correlation between intrinsic dispersion in [X/H] and structure length.  For each GALAH-sampled element, we draw 10,000 instantiations of each structure's length and intrinsic dispersion in [X/H] from normal distributions centered on the reported value and with standard deviation determined by the reported value uncertainty.  For every instantiation, we compute the Spearman Rho coefficient and its associated p-value.  When the p-value of an instantiation is less than or equal to 0.050, we consider the instantiation to contain a statistically significant correlation between length and intrinsic dispersion in [X/H].  Almost all elements have statistically insignificant correlations between structure length and intrinsic dispersion in [X/H] 99\% of the time.  K and Ba, however, are exceptions to this.  We find a statistically significant correlation between structure length and intrinsic dispersion in [K/H] and [Ba/H] 8\% and 50\% of the time, respectively.  We are suspicious of the correlation between $\rm \sigma_{[Ba/H]}$ and length because 1) determining reliable Ba abundances can be difficult, and 2) we see no such correlation present for Y, the other s-process element that we investigate in our sample.  Regarding point 1), \citet{GALAHDR3} and \citet{Spina2021}, for example, note that Ba abundances are derived from lines that are sensitive to stellar activity, and they also display a temperature dependence (Figure 16 of \citealp{GALAHDR3}).  

We explore whether the correlation between intrinsic dispersion in [Ba/H] and length is a second-order effect due to a correlation between [Ba/H] and $\rm T_{eff}$, $\rm \sigma_{[Ba/H]}$ and spread in sampled $\rm T_{eff}$, $\rm \sigma_{[Ba/H]}$ and signal to noise ratio, or $\rm \sigma_{[Ba/H]}$ and age.  We also search for correlations between [Ba/H] abundance and $\rm \sigma_{[Ba/H]}$, for Ba lines can easily become saturated and affect the precision of abundance measurements (though, in theory, our MLE approach should be unaffected by this).  We search for these correlations using the same Monte-Carlo approach that we apply in searching for correlations between structure length and intrinsic dispersion in [X/H].  We find no statistically significant correlations between [Ba/H], $\rm \sigma_{Teff}$, $\rm \sigma_{[Ba/H]}$, and structure age. We do, however, notice a potential correlation between [Ba/H] and $\rm T_{eff}$ in Theias 452 and 1415, two of the longest structures in our sample, where the intrinsic dispersions of these structures appears to be dominated by a negative correlation between [Ba/H] and $\rm T_{eff}$.  Ultimately, further investigation into this potential relationship between $\rm \sigma_{[Ba/H]}$ and length should be pursued to definitively determine its presence or lack thereof.

\subsection{Youth and Age from Li Depletion and Chemical Clocks}
\begin{figure*}
    \centering
    \includegraphics[width=1.6\columnwidth]{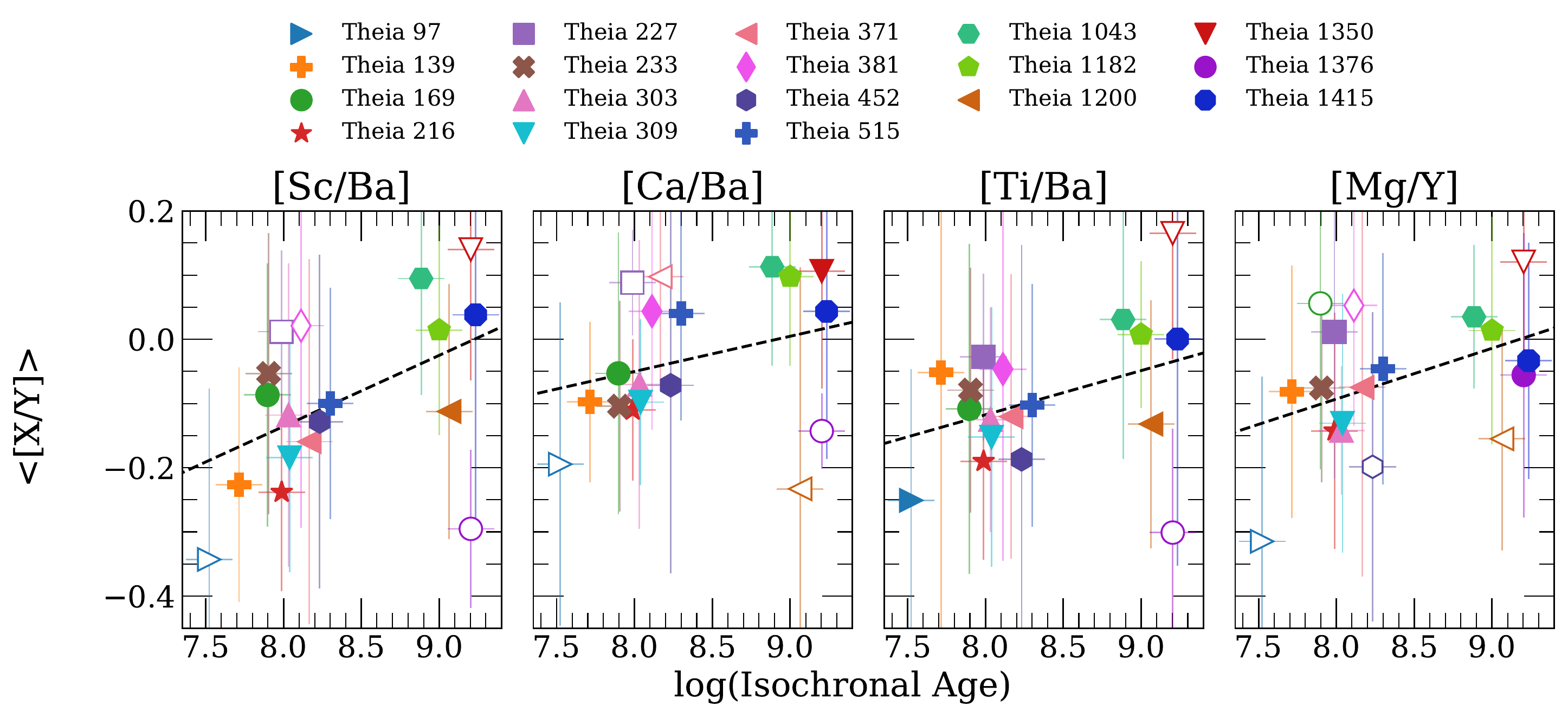}
    \caption{The mean abundance in four chemical clocks as a function of logarithmic isochronal age for each structure, with error bars representing the standard deviation in the abundance distribution.  We recreate the expected qualitative trend between age and abundance in chemical clocks that has been reported in the literature \citep[e.g.,][]{Nissen2015, Jofre2020, Hayden2020, Espinoza-Rojas2021}.  For each chemical clock, we perform a linear regression on the chemical clock abundance ratio vs. logarithmic isochronal age distribution and present the best fit line (dashed lines). Structures that lie outside of $2\sigma$ of the linear fit are considered outliers (hollowed markers).  While most structures follow the general trend between chemical clock abundance and logarithmic isochronal age, several outliers surface.  We discuss possible origins for the discrepancy between isochronal and chemical age in Section \ref{subsec:consideringage}.}
    \label{fig:chemclock}
\end{figure*}

We present the results of our A(Li) vs. T$\rm _{eff}$ analysis in Figure \ref{fig:liteff} and use these results to estimate the Li depletion ages of the structures in our sample.  To estimate ages, we compare the distributions of each structure in the A(Li)-T$\rm _{eff}$ plane with the Li tracks of clusters that have well-constrained ages. We present our age estimates in Table \ref{tab:liages} and discuss in the group-specific subsections of Appendix \ref{appendixa}.  Broadly, we find that the A(Li) vs. T$\rm _{eff}$ distributions of the groups support the isochronally-determined ages of KC19 when compared to the Li tracks of clusters with well-constrained ages.  The clearest disagreements between ages derived from Li and isochrones occur for Theia 227, which appears older than its isochronal age, and Theias 452 and 1376, which appear younger.  In addition to Li, we present the abundance ratios of four chemical clocks ([Sc/Ba], [Ca/Ba], [Ti/Ba], and [Mg/Y], see Section \ref{subsec:lithium}) as a function of logarithmic isochronal age for each structure in Figure \ref{fig:chemclock}.  We are able to recreate the expected trend between age and chemical clock abundance that has been observed by several works \citep[e.g.,][]{Nissen2015, Spina2016, Spina2018, Jofre2020, Hayden2020, Espinoza-Rojas2021}.  While most groups follow this expected trend, a few structures surface as outliers, which we define as structures with chemical clock abundance ratios that lie outside of the $2\sigma$ range of the log-linear trend created by the remaining structures.  Theia 1376 is the most extreme outlier, existing between 2.7 and 4.2$\sigma$ below the log-linear trend in all Ba-based chemical clocks.  Theias 97 and 1350 are also outliers in three of the four chemical clocks, and Theias 227, 381, and 1200 are outliers in two of the four chemical clocks.  Finally, Theias 371 and 452 are outliers in one chemical clock.  It is interesting that two of the most extreme outliers in our Li analysis (Theias 227 and 1376) are also outliers in our chemical clock analysis.  We offer potential explanations for these discrepancies between chemical and isochronal age in Section \ref{subsec:consideringage} and use this exercise to emphasize the advantages that chemistry adds to stellar age estimation.

\section{Discussion}\label{Sec:Discussion}

\subsection{Chemical Homogeneities of Strings Inform Formation Mechanisms and ISM Mixing Efficiency} \label{subsec:coevality}
In this study, we find that groups in our sample fall into three general homogeneity bins: A, groups that are just as inhomogeneous as their local fields in most elements, B, groups that are more homogeneous than their local fields but less homogeneous than OCs in most elements, and C, groups that are as homogeneous as OCs in most elements \citep[intrinsic dispersions of 0.01 to 0.07 dex, e.g.,][and references therein]{Lambert2016, Bovy2016a, Casamiquela2020, Poovelil2020}.  We find that just one structure, a string, falls into bin A, ten structures (four string-like and six compact) fall into bin B, and seven structures (five string-like and two compact) fall into bin C. We performed a Monte Carlo simulation to determine the consistency of the stellar group categorizations given the intrinsic dispersion uncertainties. We find that 15 structures remain in their classification bin in \textgreater99\% of the simulations, while the other three remain in their original classification bin in 70 - 90\% of the simulations.

It is evident that an overwhelming majority of the kinematically-identified stellar groups studied here are more chemically related than stars within their local field (Figure \ref{fig:violin1}).  This may suggest that KC19's method for identifying kinematically-related groups is also effective at identifying chemically-related groups in most cases, though further follow up should be conducted to confirm this.  Moreover, it is interesting that half of the strings in our sample, Theias 139, 169, 216, 303, and 309, are as homogeneous as OCs despite having lengths between 80 and 300 pc long (Figure \ref{fig:OCviolin2}).

What might these results suggest about the formation methods of these structures and the mixing efficiency of the molecular clouds from which they were formed?  To address these questions, we can refer to observations and simulations of the ISM and the mixing processes within molecular clouds.  Observational studies probing the MW ISM's intrinsic dispersion in elemental abundances find dispersions between 0.05 and 0.30 dex \citep[e.g.,][]{Cartledge2006, Przybilla2008, Cordova2020}, generally larger than those of OCs and associations.  This indicates that additional mixing processes within the birth molecular cloud must homogenize the gas prior to the onset of star formation.  This is supported by several theoretical studies \citep[e.g.,][]{Feng2014, Krumholz2014}.  \citet{Armillotta2018} performed adaptive-mesh three-dimensional simulations of element mixing within molecular clouds prior to star formation and conclude that turbulent mixing appears to create a characteristic homogeneity scale of 1 pc in the cloud.  That is, stars born within 1 pc of each other in a molecular cloud should be chemically homogeneous.  Additionally, stars born outside of this spatial scale are chemically homogeneous at a level that scales with the initial homogeneity of the ISM prior to the onset of cloud collapse \citep[e.g.,][]{Feng2014, Krumholz2014, Armillotta2018}.  These results suggest that the chemical homogeneity of OCs may be in part due to their being born within the small spatial scales of star forming clumps ($\rm \sim 1-4 pc$) within molecular clouds.

If we apply these inferences, which are drawn from both observational and theoretical studies, to the context of this work, we can postulate on the origins of these structures.  The low abundance dispersions of many of these strings, particularly Theias 139, 169, 216, 303, and 309, suggests that either:

\begin{enumerate}
    \item  Stars within these strings were born at small spatial scales and subsequently elongated.
    \newline OR
    \item Stars within these strings were born at large spatial scales, and the ISM can be chemically homogeneous at larger spatial scales than previously suggested.
\end{enumerate}

The first point suggests that these structures were born at the same spatial scales at which OCs are born.  This naturally leads us to question the physical mechanism(s) that could have elongated these structures to these extreme lengths (80--300 pc) within such a short timescales ($\sim$50--100 Myr, the ages of these structures).  It is possible that these highly homogeneous groups were tidally elongated during their lifetimes.  However, preliminary kinematic analysis of these structures (see \citealp{Kounkel2019} for more details) reveals that the strings are expanding radially outward from their central spine and that there is no correlation between their lengths and ages, two points of evidence that do not support this scenario.  Another explanation may involve slow and/or discontinuous star formation.  Star formation has been understood to be a rapid process that takes place over $\sim 10^5$ years \citep[e.g.,][]{McKee2007, Kruijssen2012}.  However, the theoretical work of \citet{Krumholz2018} suggests that local pockets of the ISM can retain "memory" of their chemical abundance for up to 300 Myr.  This suggests that stellar associations and OCs can be formed slowly yet still be chemically homogeneous.  It is thus possible that these structures were born of the same local pocket of the ISM but in a slow process, allowing members to disperse prior to new members forming, resulting in elongated associations.

Point (ii) suggests that these strings were formed at the long spatial scales at which they are observed.  If these structures were each born of the same elongated molecular cloud, this point could suggest that molecular clouds can be more efficient than predicted at homogenizing the ISM prior to the onset of star formation.  Alternatively, the ISM could simply be more homogeneous than previously predicted at these spatial scales.  This scenario could explain the high intrinsic dispersions in [X/H] of Theia 1415, one of the longest structures in our sample: Theia 1415 may have been born at the maximum spatial extent at which molecular clouds can effectively homogenize their contents prior to star formation, hence why it is essentially chemically indistinguishable from its local field. 

We searched for correlations between structure length and dispersion in the elements because any statistically significant correlations may support Point (ii), and any definitive lack of correlation may support Point (i).  In searching for statistically-significant correlations between length and intrinsic dispersion in [X/H], we paid particular attention to the s-process elements.  S-process elements like Ba and Y are primarily produced in and dispersed by AGB stars \citep[e.g.,][]{Karakas2014}, and theoretical works studying the correlation lengths of different elements in the ISM have found that elements ejected by AGB stars are correlated (homogeneous) on shorter spatial scales (< a few hundred pc) than those ejected in SNe \citep[> 1 kpc, e.g.,][]{Krumholz2018}.  It thus follows that s-process elements should be more sensitive tracers of birth radius than elements such as the $\alpha$-, light, and Fe-peak elements that are primarily produced in and expelled by core-collapse supernovae.  In other words, stars born together within small spatial volumes should be homogeneous in s-process elements, and this homogeneity level should scale sensitively with birth radius \citep[e.g.][]{Armillotta2018, Krumholz2018}.  A correlation between intrinsic dispersion in s-process elements and structure length would support that these structures are primordial in length.  A definitive lack of correlation, on the other hand, may support that the structures were elongated after the cessation of star formation.  Because stars trap the chemical composition of the ISM at the time and place of their birth and largely carry it with them for most of their lives \citep[e.g.][]{Feng2014}, the tidal elongation of these structures after the cessation of star formation should not affect the homogeneity levels of these structures.

As mentioned in Section \ref{Sec:Results}, we do not find any correlations between intrinsic dispersion in [X/H] and length in our sample, with the potential exception of Ba, though we suspect that this may be a second-order effect due to a potential systematic relationship between [Ba/H] and $\rm T_{eff}$ \citep[e.g., Figure 16 in][]{GALAHDR3}.  This is further supported by the lack of correlation between intrinsic dispersion in [Y/H], the other s-process element that we investigate, and length.  For the reasons outlined in the paragraph above, a definitive determination of a correlation between intrinsic dispersion in s-process elements and length, or lack thereof, would provide insight into the elongation mechanisms of these structures.  Thus, spectroscopic follow-up of these structures with an increased sample size and abundance precision and particular focus on the s-process elements should be pursued.  Though GALAH reports abundances in up to eight s-process elements (Y, Ba, La, Rb, Mo, Ru, Nd, and Sm), the structures in our analysis are only well-sampled (see Section \ref{subsec:GALAHDR3}) by GALAH in Ba and Y, so our analysis is limited to these two s-process elements.

In addition to spectroscopic followup, the two potential explanations for the elongation mechanisms and birth histories of these structures can be further probed through dynamical and gyrochronological follow up.  Dynamical analysis can be used to determine whether these strings were born compact and subsequently elongated or born in their observed filamentary shapes.  A gyrochronological study can search for age gradients in these structures to determine whether their star formation history was rapid or slow \citep[e.g.,][]{Barnes2007, Getman2018}.  Because these strings are well-sampled by TESS \citep{TESS}, gyrochronology is a natural and interesting expansion of this work.

\subsection{Connection to Giant Molecular Filaments and Milky Way Bones}
As mentioned in Section \ref{sec:intro}, stars are most commonly born in groups within molecular clouds.  While most molecular clouds are understood to be around a few tens of pcs in diameter,  recent studies have found that some molecular clouds can be between several tens and hundreds of parsecs long and just a fraction of a parsec to a few parsecs wide \citep[e.g.,][]{Ragan2014, Li2016, Soler2020}, the same dimensions as the stellar strings investigated in this study.  These elongated molecular clouds are termed giant molecular filaments, and many have been found to trace the spiral arms of the MW \citep{Goodman2014, Zucker2015}.  Theoretical studies suggest that gravitational shearing and gas compression of molecular clouds trapped down stream of spiral arms can create these elongated filamentary clouds \citep[e.g.,][]{Zucker2018}.  The spatial extents of the strings studied in this analysis, paired with the fact that older strings tend to be more diffuse and the potential correlation between intrinsic dispersion in [Ba/H] and length, suggests that these strings may have been born of giant molecular filaments.  We note that the youngest structures in our sample (such as Theias 97, 139, 169, and 515) lie along the Local Arm of the MW, while the older structures lie in the gap between the Local and Sagittarius-Carina arms.  This supports the notion that the older structures originate from giant molecular filaments that once lined spiral arms that have since shifted.  Meanwhile, the younger structures still reside in their birth spiral arm.

\subsection{Considering Stellar Chemistry in Age Estimation} \label{subsec:consideringage}
In most cases, the Li depletion patterns and abundances of chemical clocks of these groups generally support the isochronal ages determined by KC19.  There are a few exceptions, however. Theias 1376 display chemical patterns in the A(Li) vs. $\rm T_{eff}$ (Figure \ref{fig:liteff}) and chemical clock [X/Y] vs. isochronal age (Figure \ref{fig:chemclock}) planes that suggests an age younger than that determined through isochronal fitting.  Furthermore, Theia 227 appears to be older than its isochronal age according to its Li depletion pattern and relative chemical clock abundances.  Other structures, such as Theias 97, 371, 381, 452, 1200, and 1350 also show signs of a disagreement between their chemical and isochronal ages as indicated by their Li and/or chemical clock abundances.  One possible explanation for the disagreement between isochronal ages and Li depletion ages could be due to the effect that rapid stellar rotation has on suppressing Li depletion \citep[e.g.,][]{Constantino2021}.  This may particularly affect Theia 452, for the mean $vsini$ of its GALAH-sampled members is 15 $\rm km s^{-1}$, with some members displaying $vsini$ up to 25 $\rm km s^{-1}$.  The potential systematic dependencies of Ba abundances \citep[e.g.,][]{GALAHDR3, Spina2021} may also be contributing to a discrepancy between ages derived from chemical clocks and isochrones.  However, neither of these explanations would apply in cases where both Li depletion and chemical clock abundances disagree with the isochronal age in a consistent way.  In cases where both Li and chemical clock abundances tell a similar story, a possible explanation for the discrepancy between chemical and isochronal age could lie in uncertain assumptions regarding interstellar redenning, which could cause isochronal fits to under- or over-estimate structure ages.  Finally, there is a chance that we may be sampling field contaminants, rather than true structure members, in our chemical analysis.  Because field stars should generally be older, this could explain the structures with chemical ages that are older than their isochronal ages.  Spectroscopic follow-up of these structures with greater stellar sampling can minimize the effects of potential contaminants.

This work illustrates the power of considering chemistry in the study of stellar ages.  For example, in the case of isochronal age estimation, which relies on a detailed understanding of the morphology and distribution of dust in the MW, chemical indicators of age can assist in validating isochronal results.  Adding alternative age estimation methods that involve chemistry can work to either support or contest ages determined through other methods.  Calibrating chemical clocks in an absolute sense will further enhance their utility and is a natural avenue of future work in the field of stellar age.  The strings studied in this work are well-sampled by TESS \citep{TESS} and thus present an excellent avenue for calibrating chemical clocks with the more precise age determination method of gyrochronology.  Ultimately, chemical clocks appear to be a promising avenue of age estimation, and future work should seek to calibrate chemical clocks such that they can be independent indicators of age rather than relative ones.

\section{Conclusions}\label{sec:conclusion}
The recent discovery by \citet{Kounkel2019} of nearly 300 elongated stellar groups (called `strings') in Gaia DR2 presents an excellent opportunity to study the structure of the local thin disk and the star formation processes that populate it.  In this work, we examine the chemical distributions of 10 newfound stellar strings and 8 newfound non-string-like, compact stellar groups discovered by \citet{Kounkel2019} to address 1) to what degree these kinematically-related structures are chemically-related, 2) what chemistry may reveal about the formation mechanisms and birth environments of the strings in our sample.  To address these questions, we use GALAH DR3 to extract the chemical information of each structure.  We derive and report the intrinsic dispersions in elemental abundance in 17 elements for each structure and compare them to those of their surrounding local fields and to that of M 67, a well-studied open cluster.  Where possible, we validate or contest the isochronal ages of each structure using Li and other chemical clocks.

We find that all but one structure (Theia 1415) are more homogeneous than their local field in [X/H] in almost all elements by between 1.5 and 30 times.  In other words, almost all structures are not only kinematically related, a requirement for their initial discovery, but also chemically related to some degree.  Several structures, such as Theias 139, 169, 216, 227, 303, 309, 515, and 1376, are as homogeneous as open clusters (0.01 to 0.07 dex) in many elements.  We find that most structures have Li depletion patterns that support ages derived by isochronal fitting, with a few exceptions, namely Theias 452 and 1376, which appear younger than their isochronal ages, and Theia 227, which appears older.  The structures in our sample generally recreate the expected relationship between isochronal age and chemical clock abundance of [Sc/Ba], [Ca/Ba], [Ti/Ba], and [Mg/Y].  Some structures fall outside of this trend, however, and we notice that several structures with Li depletion patterns that strongly disagree with isochronal age also tend to lie off of the expected trend between chemical clock abundance and isochronal age, suggesting a discrepancy between the ages derived from chemistry and those derived from isochrones.

We conclude that chemistry is an invaluable addition to the study of stellar groups, particularly in determining population ages.  In this work, the abundances of both Li and various chemical clocks ratios work in tandem to probe structure youth and age.  We note that the absolute calibration of these chemical age/youth indicators would greatly benefit the community.  The strings studied here are well-sampled by TESS and thus would provide a useful starting point for such a calibration using gyrochronological ages.

We also conclude that the low dispersions in [X/H] found in many of the strings in our sample, paired with their long spatial extents (80-300 pc), suggests either one of two possibilities: (i) the strings were born at small spatial extents and subsequently tidally elongated, or (ii) the strings are primordial in shape.  The former scenario opens up questions of the physical mechanisms that can elongate strings to such extremes at these short (few hundred Myr) timescales.  The latter scenario suggests that either the ISM is more chemically homogeneous than previously understood or that turbulent mixing in molecular clouds is capable of homogenizing the ISM efficiently at long spatial scales.  We observe no clear correlations between intrinsic dispersion in [X/H] and length with the exception of Ba, but we suspect that this correlation is a second-order effect due to a potential systematic trend between [Ba/H] and $\rm T_{eff}$.  Further follow up with increased abundance precision and greater sampling of s-process elements should be conducted to definitively determine the presence (or lack) of a correlation between intrinsic dispersion and length.  This work highlights the unique advantages that chemistry brings into the study of kinematically-related stellar groups, providing insight into the birth histories and ages of these stellar populations in a way that can not be achieved using just kinematic and dynamical analysis.

\section*{Acknowledgements}
We thank the referee and scientific editor for helpful comments and suggestions that improved the quality of our analysis and writing.  We thank Jeff Andrews for interesting and helpful discussions that improved the quality of this work.  We thank Catherine Zucker for insightful discussions on the connection between stellar strings and molecular filaments.  We thank Kendall Sullivan and Tyler Nelson for helpful feedback that improved the clarity of and conclusions drawn by this work.

CM \& KH acknowledge support from the National Science Foundation grant AST-1907417 and AST-2108736. KH is partially supported through the Wootton Center for Astrophysical Plasma Properties funded under the United States Department of Energy collaborative agreement DE-NA0003843.

The following software and programming languages made this research possible: topcat (version 4.4; \citealt{TOPCAT}); Python (version 3.7) and its packages astropy (version 2.0; \citealt{Astropy}), scipy \citep{scipy}, matplotlib \citep{matplotlib}, pandas (version 0.20.2; \citealt{pandas}) and  NumPy \citep{numpy}. This research has made use of the VizieR catalog access tool, CDS, Strasbourg, France. The original description of the VizieR service was published in A\&AS 143, 23.

\section*{Data Availability}
This work has made use of data from the European Space Agency (ESA) mission Gaia (\url{https://www.cosmos. esa.int/gaia}), processed by the Gaia Data Processing and Analysis Consortium (DPAC, \url{https://www.cosmos.esa.int/web/gaia/dpac/consortium}). Funding for the DPAC has been provided by national institutions, in particular the institutions participating in the Gaia Multilateral Agreement.

This work has also made use of GALAH DR3, based on data acquired through the Australian Astronomical Observatory, under programmes: A/2013B/13 (The GALAH pilot survey); A/2014A/25, A/2015A/19, A2017A/18 (The GALAH survey). We acknowledge the traditional owners of the land on which the AAT stands, the Gamilaraay people, and pay our respects to elders past and present.  The GALAH DR3 data underlying this work are available in the Data Central at \url{https://cloud.datacentral.org.au/teamdata/ GALAH/public/GALAH_DR3/} and can be accessed with the unique identifier \code{galah$\_$dr3} for this release and \code{sobject$\_$id} for each spectrum.


\bibliographystyle{mnras}
\bibliography{main.bib} 

\appendix
\section{Structure-Specific Characteristics}\label{appendixa}
\subsection{Theia 97}\label{specificdisps}
Theia 97 is a young string (33 Myr according to the isochronal fits of KC19) with an aspect ratio of 4:1 and a length of 193 pc.  It displays high Li abundances (for example, A(Li)$=\sim$2.75 at $\rm T_{eff}=5000K$) that are indicative of youth, suggesting a Li depletion age between $\sim$13 and $\sim$100 Myr that supports its isochronal age.  Theia 97's abundances in [Sc/Ba], [Ca/Ba], [Ti/Ba], and [Mg/Y] also support its youth and are among the lowest in our sample.  In terms of its intrinsic dispersion in elemental abundances, Theia 97 is much less homogeneous than most OCs \citep[e.g.,][and references therein]{Lambert2016, Bovy2016a, Casamiquela2020, Poovelil2020}, with intrinsic dispersions in chemical abundances ranging from 0.11 and 0.31 dex in most elements, with the exception of [Al/H] and [Ni/H], in which it has intrinsic dispersions of $0.06^{+0.04}_{-0.03}$ dex and $0.01^{+0.11}_{-0.01}$ dex, respectively.  

\subsection{Theia 139}
Theia 139 is another young string with an aspect ratio of 15:1, a length of 196 pc, and an isochronal age of $\sim$51 Myr.  It only has $\sim$5 samples in the A(Li) vs. T$\rm _{eff}$ plane that lie in the temperature regime where Li abundances are most sensitive to age (\textless $\sim 5700$K).  However, given the available Li abundances, Theia 139 appears to be between $\sim$100 and 625 Myr.  We thus do not see clear evidence that it is as young as its isochronal age suggests.  At 51 Myr, we would expect this structure to have $\rm A(Li)\sim$2.3 at $\rm T_{eff}=5000K$, but instead we see A(Li) values of $\sim$1.8 in that temperature regime.  Despite the potential discrepancy between its isochronal age and its Li abundances, Theia 139's chemical clock abundances show no clear indication of disagreement with its isochronal age.  We measure intrinsic dispersions in [X/H] in this structure that range in value between 0.01 (Zn) to 0.16 (Ba) dex in most elements.  In addition to being highly homogeneous in Zn, it is quite homogeneous (0.04 \textless ~$\rm \sigma_{[X/H]}$~\textless~0.06 dex) in some $\alpha$- and Fe-peak elements (Mg, Si, Sc, Ti, Cr, Mn, Fe).

\subsection{Theia 169}
Theia 169 is a string with an aspect ratio of 5:1, a length of 181 pc, and an isochronal age of $\sim$79 Myr.  Like Theia 139, it has few high-quality Li samples from GALAH, making it difficult to infer a Li depletion age.  However, the available Li data largely supports its isochronal age, suggesting a Li depletion age of $\sim$100 Myr.  For all Ba-based chemical clocks, Theia 169's chemical clock abundances fall along the expected log-linear trend between isochronal age and chemical clock abundance, supporting its isochronal age.  However, this structure shows an enhancement in [Mg/Y] that exceeds $\rm 2 \sigma$ of the log-linear trend created by the remaining structures, potentially suggesting a slightly older age.  Our MLE analysis returns dispersions in [X/H] that range from 0.04 (Al) to 0.22 dex (Ba).  Theia 169 is moderately homogeneous in several light, alpha, and iron-peak elements, with intrinsic dispersions in Na, Mg, Si, Ca, Ti, Ni, and Zn between 0.05 and 0.09 dex.  Despite having similar properties to Theia 139, such as age and length, this structure has intrinsic dispersions that are consistently higher than those of Theia 139 by 0.01 to 0.10 dex, with the exception of Na, Al, Ca, and Ni, in which it displays intrinsic dispersions that are lower than those of Theia 139 by between 0.03 and 0.05 dex.

\subsection{Theia 216}
Theia 216 is a string with an aspect ratio of 6:1 and a length of 88 pc.  It has an isochronal age of $\sim$97 Myr and a Li depletion pattern and chemical clock abundances that support this young age.  Theia 216 stands out as the most chemically homogeneous structure in our sample.  We measure intrinsic dispersions in [X/H] of $0.01^{+\sim0.05}_{-0.01}$ dex for Mg, Al, Si, Sc, Mn, Ni, and Zn.  Additionally, we measure intrinsic dispersions that range from 0.02 to 0.07 dex for Ca, Ti, Cr, Fe, and Cu.  This structure, like most others, is least homogeneous in the s-process elements Y and Ba, which present intrinsic dispersions of $0.10^{+0.08}_{-0.04}$ dex and $0.13^{+0.06}_{-0.02}$ dex, respectively.

\subsection{Theia 227}
Theia 227 is a compact, spherical group with an isochronal age of $\sim$97 Myr.  This structure is not well-sampled in Li by GALAH, with just 3 stars having available Li measurements.  However, the available Li data does not support its young isochronal age.  If this structure were indeed 97 Myr old, we would expect $\rm A(Li)\sim2.0$ at $\rm T_{eff}\sim4800K$, but instead the data indicates that A(Li) $\sim$1.0 at this temperature, suggesting an older age of between 150 and 625 Myr.  Theia 227's enhancements in [Sc/Ba] and [Ca/Ba] also suggest an older age than that derived from its isochrone.  Theia 227 displays intrinsic dispersions in [X/H] that range from 0.01 (Ni, Ba) to 0.10 (Na, Sc, Cu) dex for most elements.  The only exceptions to this are Al and Y, in which Theia 227 presents intrinsic dispersions in [X/H] of $0.16^{+0.10}_{-0.05}$ dex and $0.11^{+0.13}_{-0.05}$ dex, respectively.

\subsection{Theia 233}
Theia 233 is a string with an aspect ratio of 4:1, a length of 251 pc, and an isochronal age of $\sim$80 Myr.  The two stars that lie in the temperature regime where Li is most sensitive to age (T$\rm _{eff}$ \textless$\sim$ 5700K) have A(Li) $\sim$ 2.5 $\pm$ 0.2 dex, but we would expect an A(Li) $\sim$ 2.7 for a stellar group at an age of 80 Myr.  We therefore find a 1$\rm \sigma$ difference in A(Li), and additional Li measurements, or higher precision abundances, are needed to confirm that the isochronal age differs from the age inferred from Li abundances for Theia 233.  This structure's chemical clock abundances do not show clear signs of disagreement with isochronal age.  Theia 233 displays intrinsic dispersions in [X/H] that range from 0.09 (O, Mg) to 0.29 (Ba) dex in all elements.  Theia 233 appears to be one of the least homogeneous group in our sample, something that is interesting given that it is also one of the longest and thickest ($\sim$63 pc wide) strings.

\subsection{Theia 303}
Theia 303 is a 107 Myr old string with an aspect ratio of 5:1, a length of 284 pc, and a Li depletion and chemical clock pattern that largely supports its young age.  Despite having similar properties to Theia 233 (length, width, and age), it is much more homogeneous than Theia 233 in most elements.  We measure intrinsic dispersions in [X/H] that range from 0.01 to 0.04 for O, Al, Si, Ca, Sc, Ti, Cr, Ni, Cu, and Y in this structure.  The remaining elements display intrinsic dispersions that range from 0.08 (Fe) to 0.20 (Ba) dex.  These results make Theia 303 among the most homogeneous structures in our sample, despite it being one of the longest.

\subsection{Theia 309}
Theia 309 is a string with an aspect ratio of 5:1 and a length of 193 pc.  It has an isochronal age of $\sim$109 Myr, and its Li and chemical clock abundances support this young age.  In terms of intrinsic [X/H] abundance dispersions, Theia 309 is among the most homogeneous structures in our sample.  We measure intrinsic dispersions in [X/H] for this structure that range from 0.01 (O, Mg, Si, Ca, Mn, Ni, Y) to 0.09 (Cr) dex for most elements.  Only two elements fall outside of this range: K, in which Theia 309 displays an intrinsic dispersion of $0.12^{+0.07}_{-0.05}$ dex, and Cu, in which we measure an intrinsic dispersion of $0.14^{+0.08}_{-0.03}$ dex.

\subsection{Theia 371}
Theia 371 is a spherical, non-string-like structure that is the most compact in morphology of our sample, with a diameter of $\sim$7.81 pc.  It has an isochronal age of $\sim$146 Myr and a Li depletion pattern that supports this age.  However, this structure shows a $\sim$0.20 dex enhancement in chemical clock [Ca/Ba] that may suggest an older age than that indicated by its isochrone.  Despite being the most compact structure, it is also one of the least homogeneous, with intrinsic dispersions in [X/H] ranging from 0.12 (K, Fe) to 0.25 (Cu) dex in most elements.  Si, Ca, Zn, and Ba are exceptions, displaying lower intrinsic dispersions that lie between 0.06 (Zn) and 0.11 (Si) dex.  This is the only structure in our sample in which its [Ba/H] dispersion ($\rm \sigma_{[Ba/H]} = 0.09^{+0.08}_{-0.04}$ dex) is far lower than that of most other elements.

\subsection{Theia 381}
Theia 381 is a 130 Myr old spherical, non-string-like structure.  It has just one GALAH-sampled star with a reported Li abundance that lies in the $<5700$K temperature regime where Li is most sensitive to youth, so it is difficult to draw meaningful conclusions about its youth.  However, the available Li data does not support its isochronal age.  If this structure was indeed $\sim130$ Myr, we would expect to see A(Li) $\sim$2.6 dex at $\rm T_{eff}\sim$5600K, but we instead find A(Li) $\sim$2.2 dex in this temperature regime, suggesting an age between 150 and 650 Myr.  Theia 381's chemical clock abundances also support this conclusion.  Theia 381 shares similar chemical clock abundances with Theia 227, another structure that appears to be older than its isochrone suggests.  Theia 381 displays high [Sc/Ba] and [Mg/Y] abundance ratios that lies $2\sigma$ above the expected values given the general trends created by the remaining structures, suggesting an older age.  This structure is moderately homogeneous (0.05 \textless~$\rm \sigma_{[X/H]}$~\textless~0.09 dex) in some elements, such as Na, Sc, Ti, Cr, and Mn, and quite inhomogeneous (0.20 \textless~$\rm \sigma_{[X/H]}$~\textless~0.28 dex) in others, such as Al, Ni, Cu, and Ba.  We note that the relatively high ($0.20^{+0.09}_{-0.05}$ dex) intrinsic dispersion in [Ni/H] is driven by a single outlier.  The remaining elements display intrinsic dispersions in [X/H] that lie in between those two extremes of homogeneity.

\subsection{Theia 452}
Theia 452 is a crescent-shaped string with an aspect ratio of 3:1 and a length of 486 pc, making it the longest string in our sample.  It has an isochronal age of $\sim$171 Myr, though its distribution in the A(Li) vs. T$\rm _{eff}$ plane suggests that it could be slightly younger, with an age between $\sim$13 and $\sim$150 Myr.  This structure's abundance ratio in [Mg/Y] also suggests a discrepancy between its isochronal and chemical ages.  This structure is not only the longest structure in our sample but also the least homogeneous.  Most elements have intrinsic dispersions in [X/H] no less than 0.10 (Zn) dex and up to 0.30 (Ba) dex.  The exceptions to this are Al, Ni, and Y, which display intrinsic dispersions of $0.08^{+0.05}_{-0.04}$, $0.01^{+0.10}_{-0.01}$, and $0.01^{+0.11}_{-0.01}$ dex, respectively.

\subsection{Theia 515}
Theia 515 is a non-string-like, compact spherical group with an isochronal age of 200 Myr.  Though it only has one star with Li data in the temperature regime sensitive to age ($\rm T_{eff}<5700$K), the available Li data supports its youth.  Its chemical clock abundances also support its young age.  This structure is among the poorest sampled (N$=$7) structures that we investigate, so our intrinsic homogeneity estimates have large error estimates of $\sim$0.10 dex.  Despite this, we measure intrinsic dispersions in [X/H] that range from 0.01 (O, Na, Al, Si, K, Sc, Ti, Ni, and Zn) to 0.09 (Ca) dex in most elements.  As we see in most other cases, s-process elements Y and Ba display the highest intrinsic dispersions among the elements of $0.14^{+0.11}_{-0.07}$ and $0.16^{+0.10}_{-0.06}$ dex, respectively.

\subsection{Theia 1043}
Theia 1043 is a compact, spherical group with a relatively old (in the context of this study) isochronal age of 767 Myr.  Both its Li and chemical clock abundances support this age.  We measure intrinsic dispersions in [X/H] that lie between 0.10 and 0.20 dex for all elements with the exception of Si and Ca, which display intrinsic dispersions of $0.08^{+0.07}_{-0.03}$ and $0.06^{+0.07}_{-0.03}$ dex, respectively.  In the context of the other groups investigated in this work, Theia 1043 is among the least homogeneous groups that we study.

\subsection{Theia 1182}
Theia 1182 is a compact spherical group with an isochronal age of $\sim$1 Gyr and a Li depletion pattern that notably shows no Li abundances above A(Li)$\sim$1.0 below $\rm T_{eff} < 5700K$, suggesting a lack of youth in this structure that supports its isochronal age.  Theia 1182's chemical clock abundance ratios also support its old isochronal age.  We measure large intrinsic dispersions in [X/H] in this structure that range from 0.11 (Zn) to 0.27 (Y) dex in all sampled elements.  This makes Theia 1182 among the least homogeneous compact, non-string-like structures in our sample.

\subsection{Theia 1200}
Theia 1200 is a relatively old (isochronal age $\sim$1.16 Gyr) string with an aspect ratio of 1:6 and a length of $\sim$142 pc.  Though this structure only has 2 stars with Li abundances/upper limits in the temperature regime most sensitive to age ($\rm T_{eff}<5700$K), the available data may suggest an age younger than that suggested by its isochrone.  For example, we observe one star with A(Li)$\sim$1.4 at $\rm T_{eff}\sim5100K$, which points toward a younger age of \textless625 Myr.  Theia 1200's chemical clocks also suggest a younger age.  This structure displays a [Ca/Ba] abundance ratio that is $\>4\sigma$ lower than that expected given the general trend created by the remaining structures, supporting the results of the Li depletion analysis.  Theia 1200 also displays an [Mg/Y] abundance $\rm 2\sigma$ lower than expected given the general trend in this chemical clock in the remaining structures, suggesting a younger age than that indicated by its isochrone.  We measure intrinsic dispersions in [X/H] that range from 0.10 (O, Sc) to 0.18 (Ba) dex in most elements.  Na, Ca, Cu, and Y lie outside of this regime, with intrinsic dispersions that range from 0.21 to 0.35 dex for Na, Ca, and Cu, and an intrinsic dispersion of $0.04^{+0.13}_{-0.01}$ for Y.

\subsection{Theia 1350}
Theia 1350 is a compact, spherical structure with a relatively old isochronal age of 1.6 Gyr.  This structure's distribution in the A(Li)-$\rm T_{eff}$ plane generally supports this age.  By contrast, Theia 1350's abundance ratios of [Sc/Ba], [Ti/Ba], and [Mg/Y] lie 2$\sigma$, 3$\sigma$, and 2$\sigma$, respectively, above the expected value given the chemical clock-isochronal age trend created by the remaining structures, suggesting an older age.  Like Theia 1182, it is among the least homogeneous compact structures in our sample, with intrinsic dispersions in [X/H] that range from 0.12 (Si, Sc) to 0.24 (Cr, Y, Ba) dex in most elements, with the exception of O, Mg, and Al, which display intrinsic dispersions of $0.01^{+0.07}_{-0.01}$, $0.10^{+0.06}_{-0.03}$, and $0.09^{+0.06}_{-0.03}$ dex, respectively.

\subsection{Theia 1376}
Theia 1376 is the oldest compact spherical structure in our sample with an isochronal age of 1.6 Gyr.  The available Li detections seem to follow a cohesive track on the A(Li) vs. T$\rm_{eff}$ plane indicating an age between 150 and 625 Myr, much younger than its isochronal age.  For example, at $\rm T_{eff}\sim 5200K$, we observe an A(Li) of $\sim$1.7, whereas were this structure truly 1.6 Gyr, we would not expect to see any Li present in stars in that temperature regime.  This structure's chemical clock abundances tell the same story.  Theia 1376's abundances in [Sc/Ba], [Ca/Ba], and [Ti/Ba] are between 0.20 and 0.40 dex lower than expected given its isochronal age, pointing towards youth.  Interestingly, its abundance in [Mg/Y] does not suggest a clear discrepancy between its chemical and isochronal ages.  Theia 1376 is one of the most homogeneous structures in our sample.  We measure intrinsic dispersions in [X/H] between 0.01 and 0.08 dex level for Al, Si, K, Ca, Sc, Ti, Cr, Mn, Zn, Y, and Ba.  The remaining elements display intrinsic dispersions in [X/H] between 0.13 (Na, Mg) and 0.23 (Cu) dex.

\subsection{Theia 1415}
Theia 1415 is the oldest string in our sample, with an aspect ratio of 5:1 and a length of 350 pc.  It has an isochronal age of 1.72 Gyr and a lack of Li that is indicative of old age.  Theia's 1415 chemical clock abundances also support its old age.  We measure high intrinsic dispersions in [X/H] for this group that range from 0.16 (Ca) to 0.35 (Ba) dex for all elements.  It is interesting that Theia 1415 is one of the longest and thickest (72 pc) strings and also one of the most inhomogeneous in [X/H] in our sample.


\bsp	
\label{lastpage}
\end{document}